\documentclass{emulateapj}

\def\lsim{\mathrel{\rlap{\lower4pt\hbox{\hskip1pt$\sim$}}
    \raise1pt\hbox{$<$}}}                
\def\gsim{\mathrel{\rlap{\lower4pt\hbox{\hskip1pt$\sim$}}
    \raise1pt\hbox{$>$}}}                

\shorttitle{Very Deep {\it Chandra} Observation of A2052}
\shortauthors{Blanton et al.}

\slugcomment{Accepted for publication in ApJ}

\begin{document}

\title{A Very Deep {\it Chandra} Observation of Abell~2052:  Bubbles, Shocks, and Sloshing}

\author{E. L. Blanton\altaffilmark{1,2},
S. W. Randall\altaffilmark{2},
T. E. Clarke\altaffilmark{3}
C. L. Sarazin\altaffilmark{4},
B. R. McNamara\altaffilmark{5,2,6},
E. M. Douglass\altaffilmark{1},
and M. McDonald\altaffilmark{7}}

\altaffiltext{1}{Institute for Astrophysical Research and Astronomy Department,
Boston University, 725 Commonwealth Avenue, Boston, MA  02215;
eblanton@bu.edu, emdoug@bu.edu}

\altaffiltext{2}{Harvard Smithsonian Center for Astrophysics,
60 Garden Street, Cambridge, MA 02138;
srandall@head.cfa.harvard.edu; ELB as Visiting Scientist}

\altaffiltext{3}{Naval Research Laboratory, 4555 Overlook Avenue SW, 
Washington D. C. 20375; tracy.clarke@nrl.navy.mil}

\altaffiltext{4}{Department of Astronomy, University of Virginia,
P. O. Box 400325, Charlottesville, VA  22904-4325;
sarazin@virginia.edu}

\altaffiltext{5}{Department of Physics and Astronomy, University of Waterloo,
Waterloo, ON N2L 2G1, Canada;
mcnamara@sciborg.uwaterloo.ca}

\altaffiltext{6}{Perimeter Institute for Theoretical Physics, 31 Caroline St., N. Waterloo,
Ontario, Canada, N2L 2Y5}

\altaffiltext{7}{Astronomy Department, University of Maryland, College Park, MD 20742; mcdonald@astro.umd.edu}

\begin{abstract}
We present first results from a very deep ($\sim650$ ksec) {\it Chandra} X-ray observation of Abell 2052, as well
as archival VLA radio observations.  The data reveal detailed structure in the inner parts of the cluster,
including bubbles evacuated by the AGN's radio lobes, compressed bubble rims, filaments, and loops.  Two
concentric shocks are seen, and a temperature rise is measured for the innermost one.  On larger scales, we report
the first detection of an
excess surface brightness spiral
feature.  The spiral has cooler temperatures, lower entropies, and higher abundances
than its surroundings, and is likely the result of sloshing gas initiated by a previous cluster-cluster
or sub-cluster merger.  
Initial evidence
for previously unseen bubbles at larger radii related to earlier outbursts from the AGN is presented.
\end{abstract}


\section{Introduction} \label{sec:intro}
With its sub-arcsec resolution, the {\it Chandra} X-ray Observatory has 
revealed a wealth of substructure
in the X-ray-emitting intracluster medium (ICM) in clusters of galaxies.
``Cavities'' or ``bubbles'' are commonly seen in the X-ray gas in cluster
centers related to outbursts by the active galactic nucleus (AGN).
In addition, surface brightness edges associated with shocks and cold fronts
have been observed, as well as spiral features likely related to
``sloshing'' of the intracluster medium in cluster centers.

The central bubbles are frequently seen in cooling flow (or ``cool core'')
clusters and are often filled with radio emission associated with the AGN.
There was some evidence for these features in a very few cases even before 
{\it Chandra} observations (e.g.\ {\it ROSAT} observations of Perseus [B\"ohringer et al.\ 1993], 
Abell 4059 [Huang \& Sarazin 1998], and Abell 2052 [Rizza et al.\ 2000]).  
Clusters with cool cores have radio bubbles much more often than non-cool core
clusters (e.g. Mittal et al.\ 2009), and some of the most spectacular
individual cases observed by {\it Chandra} are the Perseus cluster
(Fabian et al.\ 2003, 2006),
M87/Virgo (Forman et al.\ 2005, 2007),
Hydra A (McNamara et al.\ 2000, Wise et al.\ 2007),
MS0735.6+7421 (McNamara et al.\ 2005), 
and Abell 2052 (Blanton et al.\ 2001, 2003, 2009, 2010).

The classic ``cooling flow problem'' is that sufficient quantities of cool
gas (as evidenced by star formation rates, for example) were not found to
match the gas cooling rates estimated from earlier X-ray observatories.
The X-ray cooling rates have lowered based on {\it Chandra} and {\it XMM-Newton}
observations, and the majority of the gas is seen to cool to only some
fraction (typically one-third to one-half) of the cluster average temperature (Peterson et al.\ 2003).
A heating mechanism is required to stop the gas from cooling to even lower temperatures.

The most likely candidate for heating of the central cluster gas is feedback
from central AGN that are fed by the cooling gas (see McNamara \& Nulsen 2007 for a review).  
This heating comes in the form
of bubbles inflated by the AGN, described above, that then rise buoyantly to larger
radii in cluster atmospheres distributing the heat.  The details of this heating, 
however, are still not completely understood.  In addition, there is at least
some component of shock heating in many cluster centers, especially early on in
the AGN's lifetime.  Shocks have been detected in fewer cases than radio bubbles,
and observations include those of Perseus (Fabian et al.\ 2003, 2006; Graham et al.\ 2008),
M87/Virgo (Forman et al.\ 2005), Hydra A (Nulsen et 
al.\ 2005a), Hercules A (Nulsen et al.\ 2005b), MS0735.6+7421 (McNamara et 
al.\ 2005), and the group NGC 5813 (Randall et al.\ 2011).
Typically, the shocks are weak with Mach numbers ranging from approximately
1.2 to 1.7 (McNamara \& Nulsen 2007).  In even fewer cases are temperature rises
detected associated with the shocks.

Heating of the ICM may also result from gas ``sloshing'' 
in the cluster's central potential well (Ascasibar \& Markevitch 2006; ZuHone et al.\
2010).  Cold fronts, regions where the surface brightness changes sharply
but the pressure does not, can result from these sloshing motions driven initially
by cluster or sub-cluster mergers.  The sloshing can produce a spiral distribution
of cool cluster gas reaching into the cluster center (e.g.\ Clarke et al.\ 2004,
Lagan\'a et al.\ 2010), and is also seen in galaxy groups (Randall et al.\ 2009a).

Here, we present a very deep {\it Chandra} observation of the cool core cluster
Abell 2052.  The only other cool core clusters observed to similar or greater depth with {\it Chandra}
are Perseus (Fabian et al.\ 2006) and M87/Virgo (Million et al.\ 2010, Werner et al.\ 2010).
Abell~2052 is a moderately rich cluster at a redshift of
$z=0.03549$ (Oegerle \& Hill 2001).
Its central radio source, 3C~317, is hosted by the central cD galaxy, UGC 09799.
Abell~2052 was previously observed in the X-ray with $Einstein$
(White, Jones, \& Forman 1997),
{\it ROSAT}
(Peres et al.\ 1998, Rizza et al.\ 2000),
{\it ASCA}
(White 2000), 
{\it Chandra} (Blanton et al.\ 2001, 2003, 2009, 2010),
{\it Suzaku}
(Tamura et al.\ 2008),
and {\it XMM-Newton}
(de Plaa et al.\ 2010).
In addition,
we present archival radio observations from the Very Large Array (VLA).


We assume $H_{\circ}=70$ km s$^{-1}$ Mpc$^{-1}$, $\Omega_{M}=0.3$, and 
$\Omega_{\Lambda}=0.7$ ($1\arcsec = 0.7059$ kpc at $z = 0.03549$) throughout.  Errors
are given at the $1\sigma$ level unless otherwise stated.

\section{Chandra Observations and Data Reduction} \label{sec:data}

Abell~2052 was observed with {\it Chandra} for a total of 662 ksec in Cycles 1, 6, 
and 10 from 2000 -- 2009.  A summary of the observations is given in Table 1.
All observations were performed with the ACIS-S as the primary instrument.  The ACIS-S
was chosen for its greater response at low energies than the ACIS-I, yielding a higher
count rate for this relatively cool, $kT\approx3$ keV cluster.  The nominal roll angles used for the
observations varied from $99^{\circ}$ to $265^{\circ}$ providing coverage at large radii
at a range of azimuthal angles around the cluster.
The events were telemetered in Very Faint mode for all but the Cycle 1 observation, where
the events were telemetered in Faint mode.
The data were processed in the standard manner, using CIAO 4.2 and CALDB 4.2.2.
After cleaning, and filtering for background flares (using the 2.5 -- 7.0 keV range for the BI chips and the 
0.3 -- 12.0 keV range for the FI chips), the total exposure remaining for the eleven data sets was
657 ksec.  Background corrections were made using the blank-sky background fields, including the
new ``period E'' background files for the more recent data.  For each target events file, a corresponding
background events file was created, normalizing by the ratio of counts in the 10 -- 12 keV energy range 
for the source and background files to set the scaling.

\section{Radio Data} \label{sec:radio}

We have used the NRAO data achive to extract observations of 3C 317 at
4.8 and 1.4 GHz. A summary of the data sets is presented in
Table~\ref{tbl:radio}. The archival radio data were calibrated and
reduced with the NRAO Astronomical Image Processing System
(AIPS). Images were produced through the standard Fourier transform
deconvolution method for each frequency and configuration. Several
loops of imaging and self-calibration were undertaken for each data set
to reduce the effects of phase and amplitude errors in the data. The
final radio image at 4.8 GHz was obtained through combining the three
VLA configurations and two observing frequencies. We have also
produced a combined configuration image at 1.4 GHz but do not show it as the 
structure is remarkably similar to the 4.8 GHz image.

We have made a spectral index map between 1.4 GHz and 4.8 GHz to
compare the spectral features more directly with the X-ray
structure. This map was created from the combined configuration data
at each frequency. The $uv$-coverage of both data sets was matched and
both frequencies were imaged with a 4.3\arcsec\ circular beam. We have
blanked all pixels in the spectral index map that were lower than the
5$\sigma$ level on either of the input maps. 

\section{Images} \label{sec:image}

\begin{figure}
\plotone{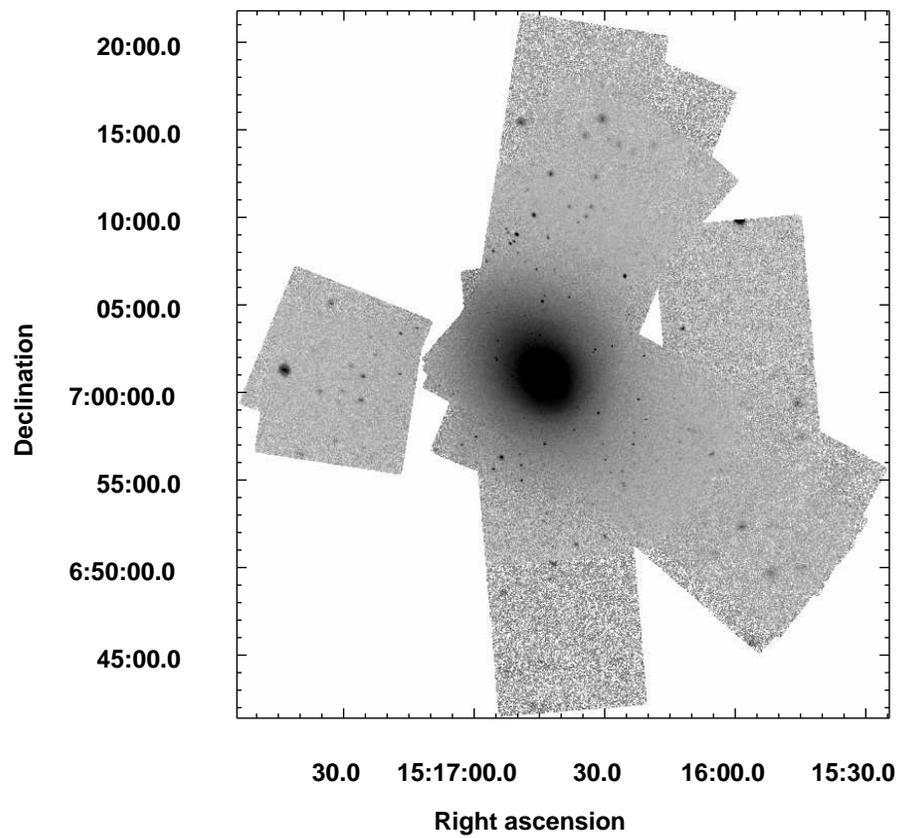}
\caption{{\it Chandra} ACIS image of A2052 in the $0.3 - 10.0$ keV band showing all CCD chips used in the analysis
(the S4 chip was excluded) and illustrating
the different roll angles used in the observations.
\label{fig:allchips}}
\end{figure}

Merged X-ray images of the source were created.  The images were corrected using merged background images and
exposure maps.  An image showing the combined data from all CCD chips used in the analysis (ACIS S1, S2, S3, I2, and I3) 
in the $0.3 - 10.0$ keV band
is shown in Fig.\ \ref{fig:allchips}.  This figure illustrates the different roll angles that were
used when the observations were performed.  The extended cluster emission is visible, as well as numerous
point sources which appear more extended while increasingly off-axis due to the larger PSF in these
regions.  The image has been smoothed with a 3\arcsec\ Gaussian.

\begin{figure}
\plotone{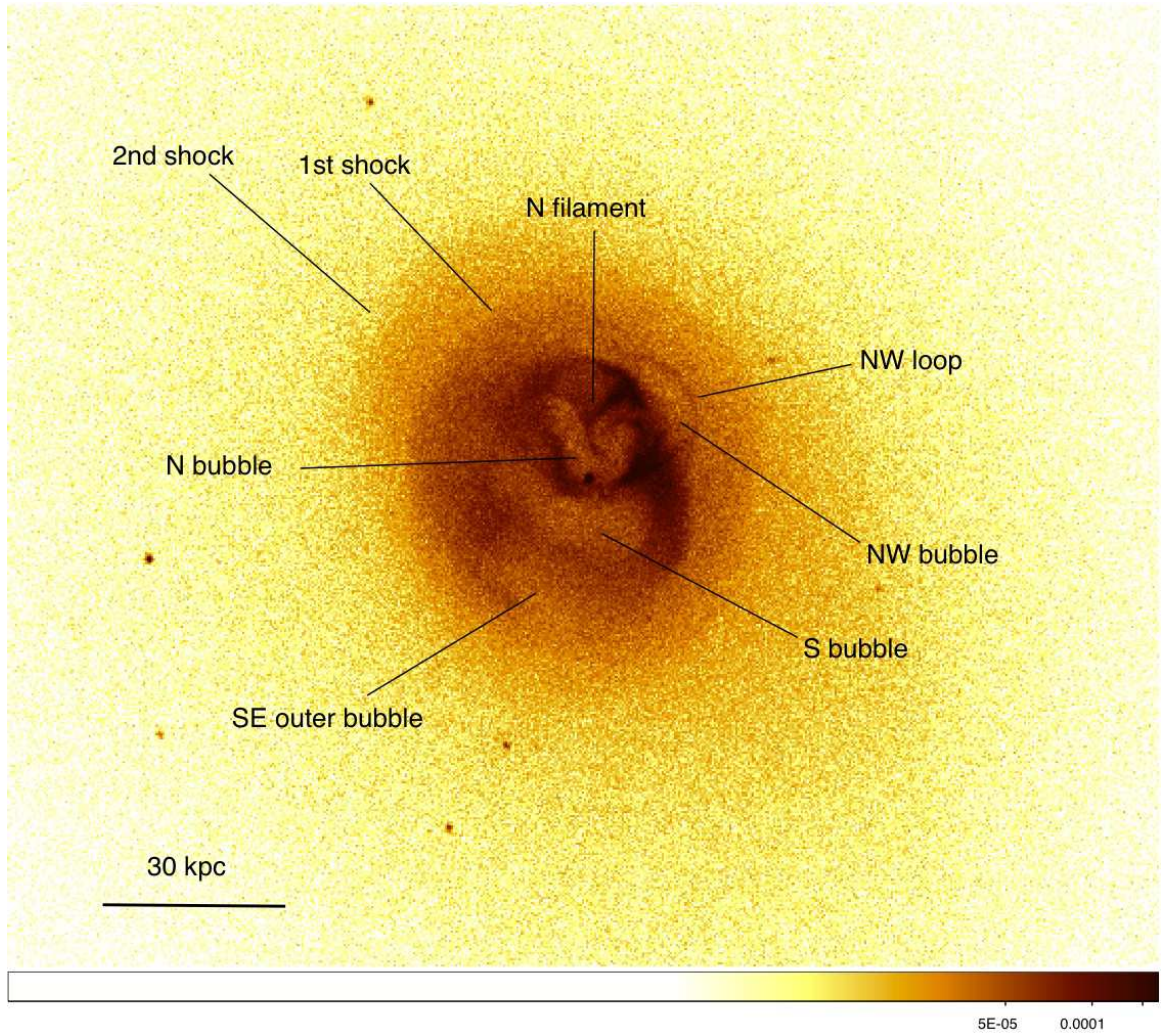}
\caption{Unsmoothed {\it Chandra} image in the $0.3 - 2.0$ keV band of the central region of A2052.  The image reveals detailed
structure related to the interaction of the AGN with the ICM, with important features labeled.
\label{fig:unsmoo}}
\end{figure}

An unsmoothed image in the $0.3 - 2.0$ keV band of the central region of the cluster is shown in Fig.\ \ref{fig:unsmoo}.
The image reveals exquisite detail related to the interaction of the AGN with the ICM.  Point sources, including the
central AGN, are visible.  Cavities or bubbles to the N and S of the AGN are seen, as well as outer cavities to the
NW and SE.  The inner cavities are bounded by bright, dense, rims, and the NW cavity is surrounded by a very narrow, 
filamentary loop.  A filament extends into the N bubble.  A shock is seen exterior to the bubbles and rims, and a probable
second shock is visible to the NE.

A three-color image of the central $6\farcm56 \times 6\farcm56$ region of A2052 is displayed in Fig.\
\ref{fig:3color}.  The image has been slightly smoothed, using a 1\farcs5 Gaussian, and the scaling for
each of the three colors is logarithmic.  Red represents the soft ($0.3 - 1.0$ keV) band, green is medium
($1.0 - 3.0$ keV), and blue represents the hard band ($3.0 - 10.0$ keV).  
The bright rims surrounding the inner bubbles appear cool.
The first shock exterior to the bubble rims is slightly elliptical with the major axis in
the north-south direction (Blanton et al.\ 2009).  It is visible as a discontinuity in the surface brightness, and the bluish color
is consistent with a temperature rise in this region.  Outside of the inner shock, a second surface brightness discontinuity
is seen as greenish emission in this figure.  The discontinuity is sharper to the NE and more extended to the SW.  This
feature may represent a shock or a cold front.  These features will be explored in further detail in \S\ref{sec:shocks}.

\begin{figure}
\plotone{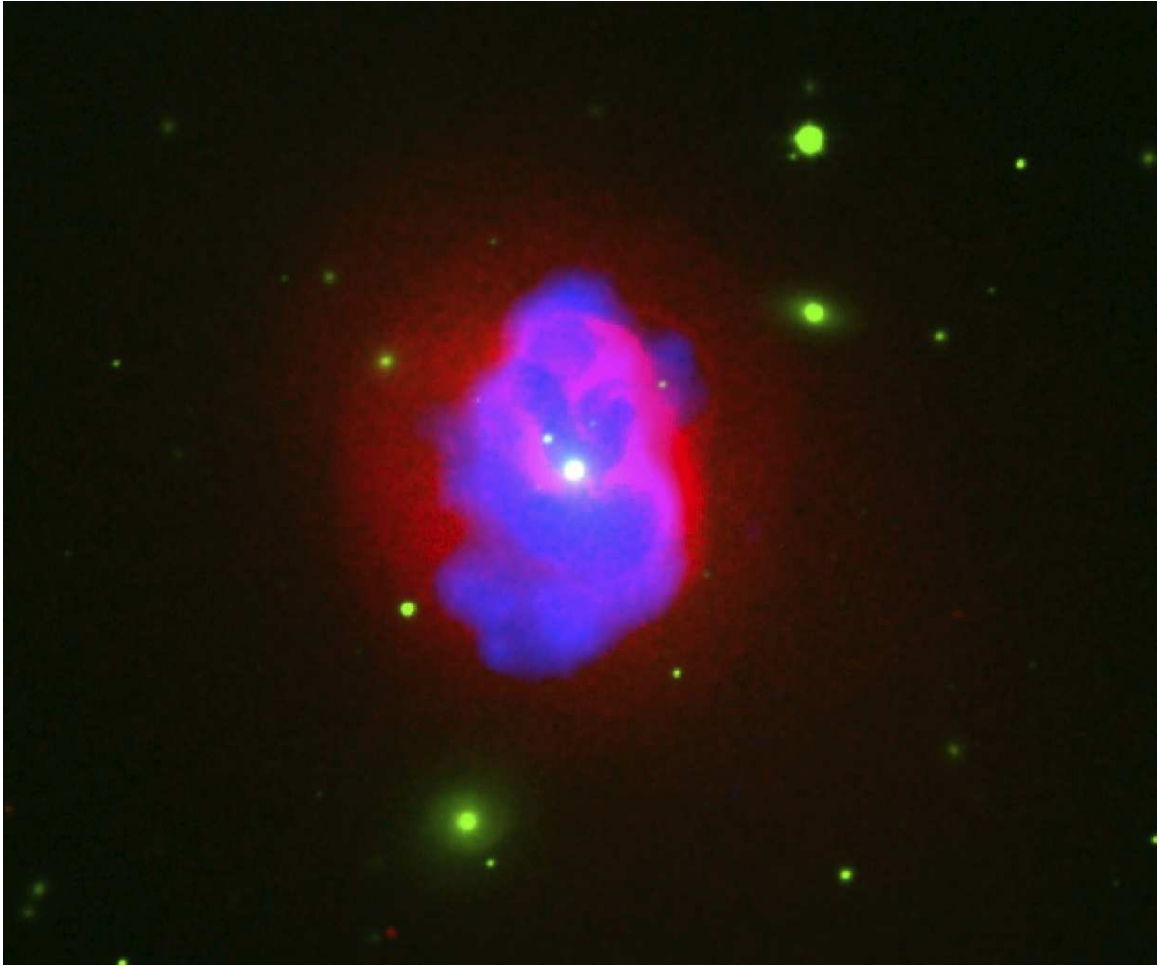}
\caption{Composite {\it Chandra} X-ray (red), VLA 4.8 GHz (blue), and SDSS r-band (green) 
$6\farcm6 \times 5\farcm8$ image of Abell~2052.
\label{fig:composite1}}
\end{figure}

\begin{figure}
\plotone{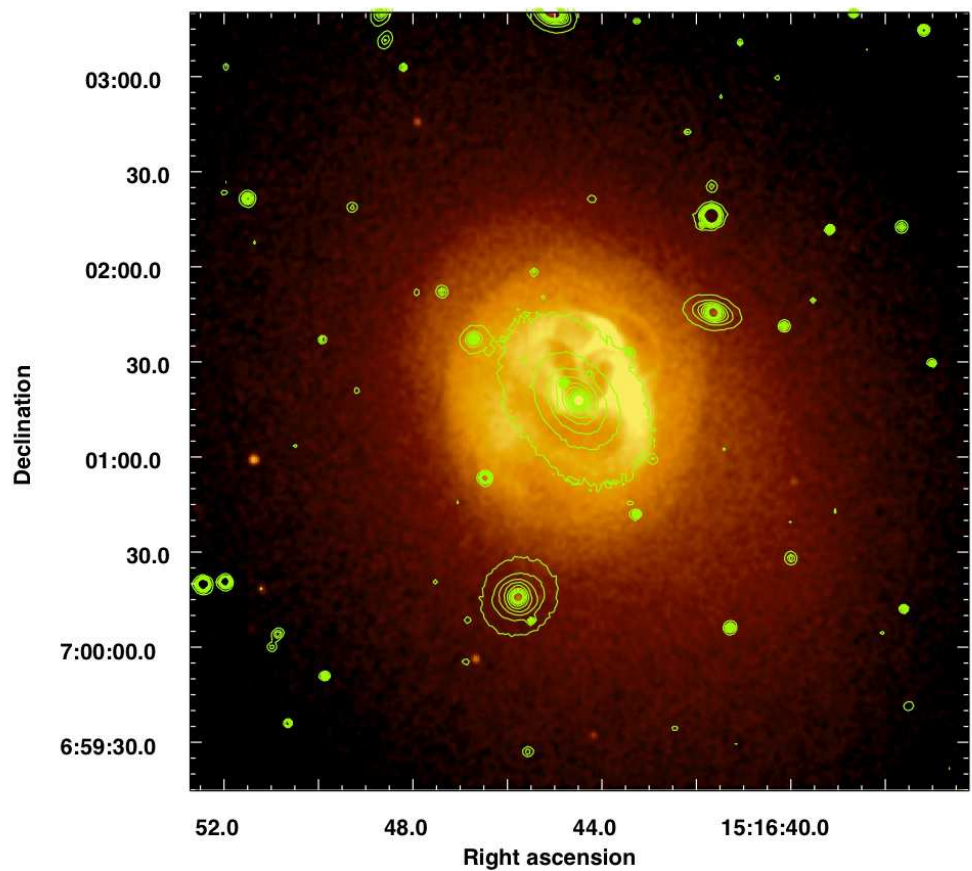}
\caption{{\it Chandra} image in the $0.3 - 10.0$ keV band, smoothed with a $1\farcs5$ radius Gaussian, with contours
of optical r-band emission from the SDSS superposed.
\label{fig:sdsscont}}
\end{figure}

A composite X-ray/optical/radio image is shown in Fig.\ \ref{fig:composite1}.
The $0.3 - 2.0$ keV {\it Chandra} image is shown in red, radio emission at 4.8 GHz from the VLA is
displayed as blue, and optical r-band emission from the Sloan Digital Sky Survey (SDSS; Abazajian et al.\ 2009) is shown as green.
The AGN is visible in the X-ray, radio, and optical, and the radio lobes fill the cavities in the X-ray emission.  This includes
the inner cavities as well as an outer cavity bounded by a narrow loop to the NW and the outer cavity to the S/SE.  In addition, the 
radio emission is breaking through the northern bubble rim to the north.  The X-ray filament extending from the northern bubble rim
towards the AGN was found to be associated with $H\alpha$ emission in Blanton et al.\ (2001).  In Fig.\ \ref{fig:sdsscont}, we
show SDSS r-band contours superposed on a {\it Chandra} image in the $0.3 - 10.0$ keV range that has been smoothed with a
$1\farcs5$ radius Gaussian.  The central cD galaxy is oriented in the NE-SW direction in the optical.  The inner bubbles seen
in the X-ray emission are within the cD galaxy.

\subsection{Residual Images} \label{sec:resids}

In order to better reveal features in the X-ray image, we created residual images using two different techniques.  In the first,
we used the method of unsharp-masking, and in the second, we subtracted a 2D beta model from the X-ray image.  We find that
unsharp-masking is useful for highlighting the structure in the inner parts of the cluster, while the model subtraction is better
at revealing larger-scale features.

\subsubsection{Unsharp-masking}

We created an unsharp-masked image in the $0.3 - 10.0$ keV energy range.  Sources were detected in the image using the wavelet detection
tool ``wavdetect'' in CIAO (Freeman et al.\ 2002).  Several wavelet scales were used, at 1, 2, 4, 8, and 16 pixels, where 1 pixel = $0\farcs492$.
Sources detected using this method were visually examined and several were rejected as being clumpy X-ray gas emission rather than
point sources.  A source-free image was created by replacing the source pixel values with the average value found in an annulus surrounding
each source.  We retained the sources in our unsharp-masked image while making corrections with a source-free image.  Smoothed images,
both with and without sources, were created by smoothing with a $0\farcs98$ radius Gaussian.  Another copy of the source-free image was smoothed with a $9\farcs8$ Gaussian.
A summed image was made by combining the source-free images smoothed at the two different scales.  A difference image was made by 
subtracting the $9\farcs8$ Gaussian-smoothed source-free image from the $0\farcs98$ Gaussian-smoothed image that contained sources.  Finally,
the unsharp-masked image was created by dividing the difference image by the summed image.  In this way, we retain the sources in the image,
smoothed at a scale of $0\farcs98$.  This method is similar to that in Fabian et al.\ (2006), although sources are 
excluded throughout in their unsharp-masked images.

\begin{figure}
\plotone{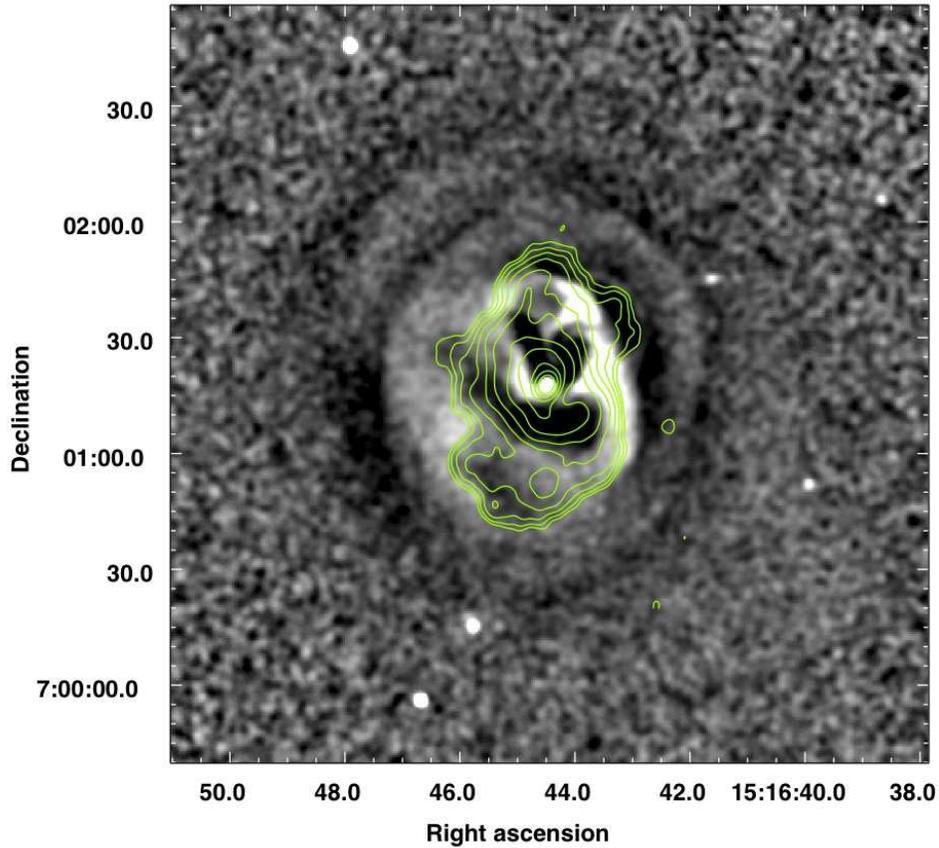}
\caption{Unsharp-masked {\it Chandra} $0.3 - 10.0$ keV image of A2052 with 4.8 GHz radio contours superposed.
Radio emission fills X-ray cavities to the N and S, as well as outer cavities to the NW, S-SE, and E.  The cavities
are surrounded by X-ray bright rims.  A filament extends into the N cavity, and a narrow filament surrounds the NW
hole.  Ripple-like features are seen surrounding the cluster center, corresponding to weak shocks.
\label{fig:unsharp}}
\end{figure}

The unsharp-masked image is displayed in Figure \ref{fig:unsharp} with VLA 4.8 GHz radio contours superposed.
The bubbles in the X-ray emission are more easily seen in this image, as are the bright bubble rims, the shock exterior to the bubble rims, and
the second shock or cold front feature to the NE.  The radio emission fills the inner bubbles and the southern lobe turns to fill the outer southern
bubble.  The radio lobe to the north appears to be escaping through a gap in the northern bubble rim, and a narrow filament is seen in the radio
in this region.  The radio emission also extends beyond the bubble rims to the NW to fill a small bubble bounded by a narrow X-ray filament.
To the east, an extension in the radio fills a small depression in the X-ray on the scale of approximately $10\arcsec$.

\subsubsection{Beta-model subtraction} \label{sec:beta}

A 2D beta model was used to fit the surface brightness in both $0.3 - 2.0$ keV and $0.3 - 10.0$ keV images using a circular region with
radius $5\farcm66$.  The images were source-free, with the surface brightness at the position of sources approximated from the surface brightness
in an annulus around each source, as above.  Corrections were made for exposure, using a merged exposure map.  Similar results were obtained 
for the fits to the images in both energy bands.  Errors were computed using Cash statistics.  For the $0.3 - 2.0$ keV image, the center 
of the large scale
emission was found to be only $1\farcs2$ away from the position of the AGN.  The emission was found to be slightly elliptical, with an
ellipticity value of $0.18\pm0.00081$ (where ellipticity values range from 0 to 1, with 0 indicating circular emission).  The position angle for
the semi-major axis of the ellipse is $38.2\pm{0.1}^{\circ}$ measured north towards east.  The core radius using this model is 
$25.1\pm{0.060}\arcsec$ and the beta index is $\beta = 0.46\pm{0.00021}$.

The residual image after 2D beta model subtraction in the $0.3 - 2.0$ keV band in shown in Fig.\ \ref{fig:beta2d}.  The image
has been smoothed with a $7\farcs38$ radius Gaussian.  The smoothing washes out the details in the very center of the image, but
the bright bubble rims are clearly visible.  A spiral feature is seen, starting in the SW and extending to the NE.
Similar spiral structures have been seen in other clusters, with A2029 being a particularly clear example (Clarke et al.\ 2004).

\section{X-ray Spectral Maps} \label{sec:maps}

Spectral maps were created to examine the distribution of temperature, as well as entropy, pressure, and abundance using the
technique described in Randall et al.\ (2008, 2009b).  The temperature
maps were created by extracting spectra for the separate ${\it Chandra}$ observations and fitting them simultaneously 
in the $0.6 - 7.0$ keV range with 
a single temperature APEC model, with $N_H$ set to the Galactic value of $2.71\times10^{20}$ cm$^{-2}$ (Dickey \& Lockman 1990)
and the abundance allowed to vary.  Background spectra
were extracted from the blank sky background observations that were reprojected to match each data set.  Data from both the frontside-
and backside-illuminated (FI and BI) chips were used, with separate response files and normalizations determined for each ObsID's data set,
with the FI and BI normalizations allowed to vary independently for each ObsID.
Spectra were extracted with a minimum of either 2000 or 10000 background-subtracted counts, corresponding to a minimum SNR of
approximately 45 or 100, respectively, depending on our analysis goals.  The radius of the circular extraction region
was allowed to grow to the size required to extract the minimum counts.
Spectral maps of the central region of A2052 were previously presented in Blanton et al.\ (2003, 2009).  Here, with the much deeper
{\it Chandra} data, we are able to examine a much larger region of the cluster with high precision.
We have chosen to include spectral maps created as described above rather than Voronoi-Tesselation maps, even though some of the map
pixels are not independent (some regions used in spectral fitting overlap), especially in the outer regions of the maps.  We find that spectral 
structures are easier to see on the maps we present, and they are confirmed by extracting spectral profiles in \S6 and \S7.

\begin{figure}
\plotone{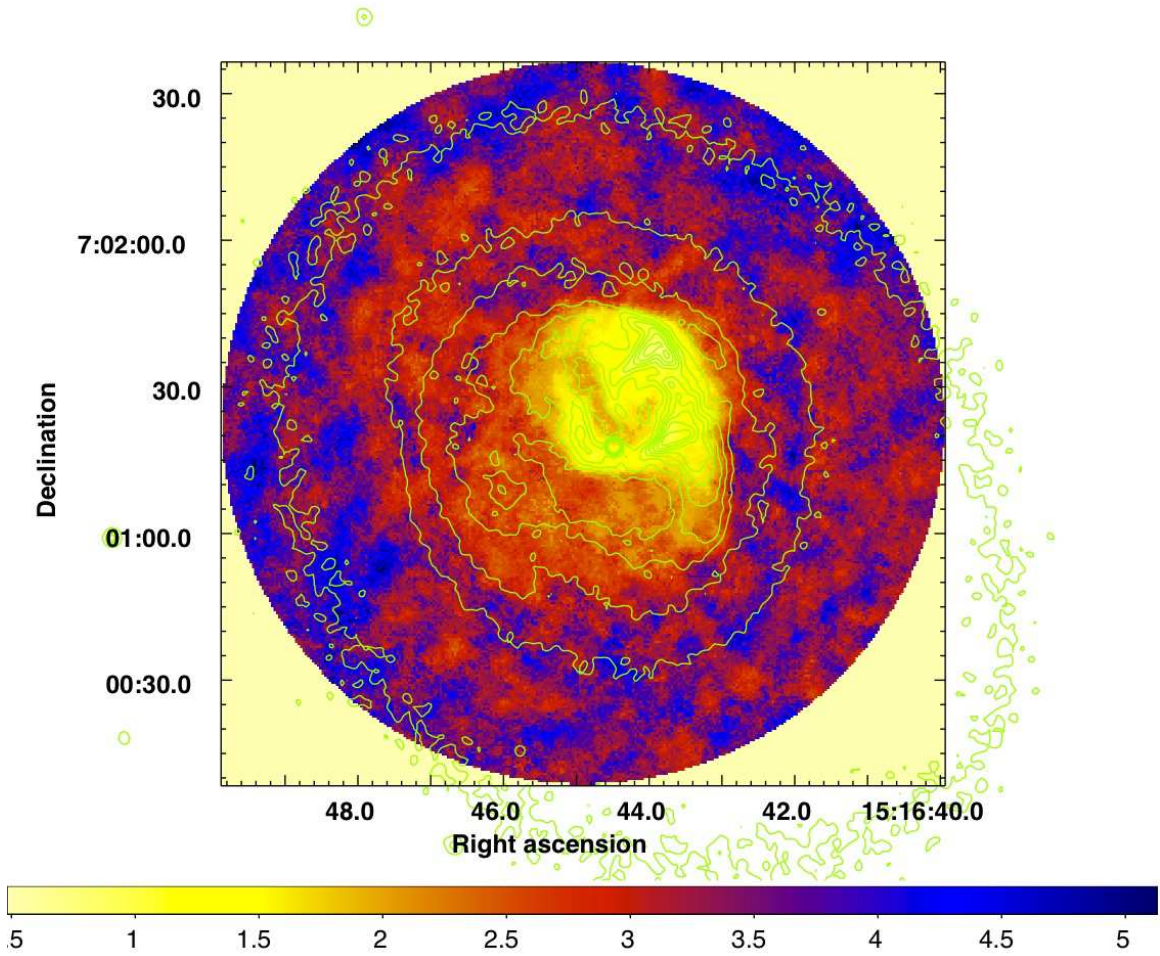}
\caption{High-resolution temperature map of the central region of A2052 with X-ray surface brightness contours in the $0.3-10.0$ keV range
superposed.  The scale bar is kT in units of keV.  The rims surrounding the X-ray cavities are cool, and the 
coolest regions, in projection, are coincident with the brightest regions of the rims.
\label{fig:Thighrescen}}
\end{figure}

\begin{figure}
\plotone{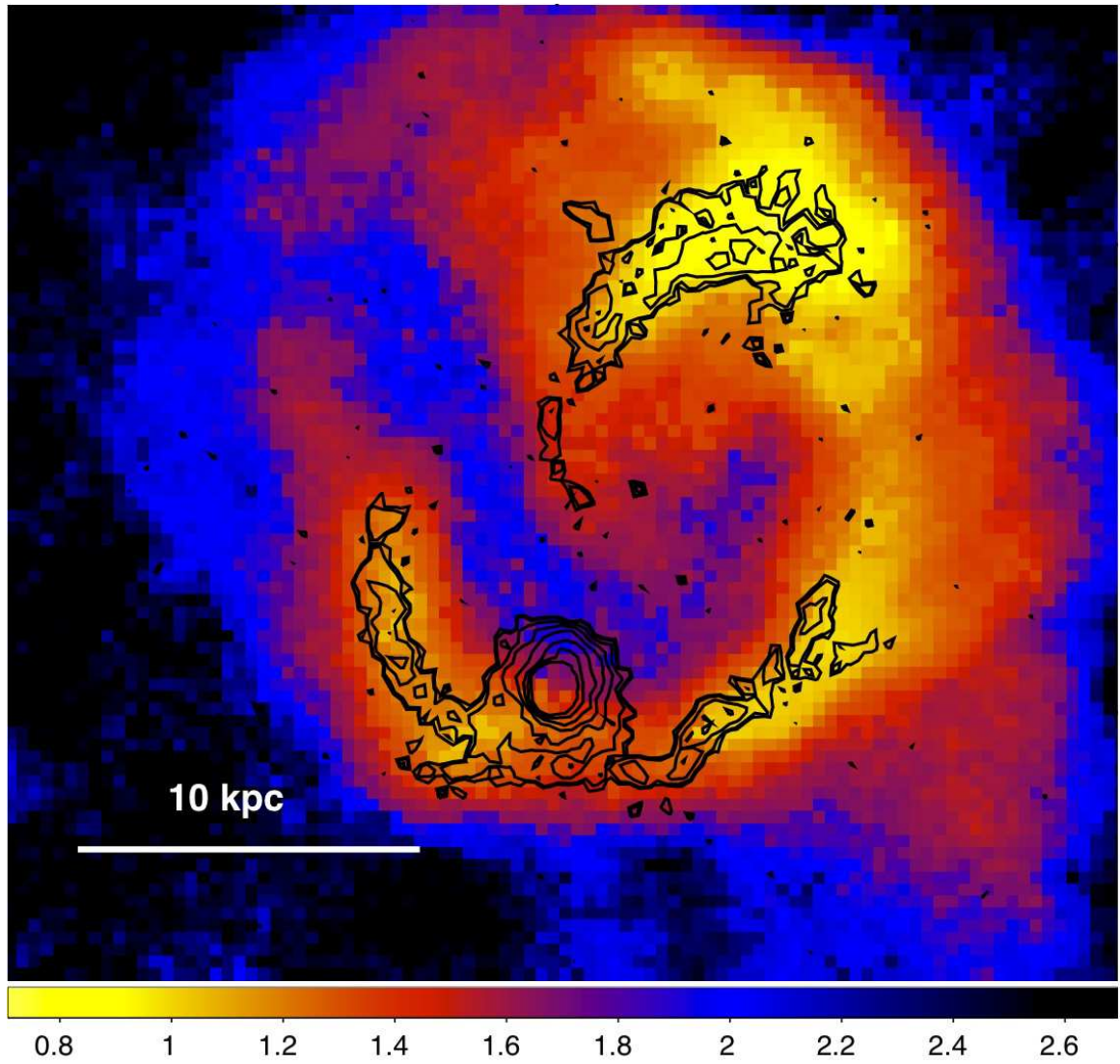}
\caption{$H\alpha$ contours from McDonald et al.\ (2010) superposed onto the central portion of the high-resolution temperature map from Fig.\ \ref{fig:Thighrescen}.
The coolest regions seen in the X-ray are coincident with emission in $H\alpha$, representing gas with $T\approx10^4$ K.  The scale bar
is kT in units of keV.
\label{fig:halphatmap}}
\end{figure}

\begin{figure}
\plotone{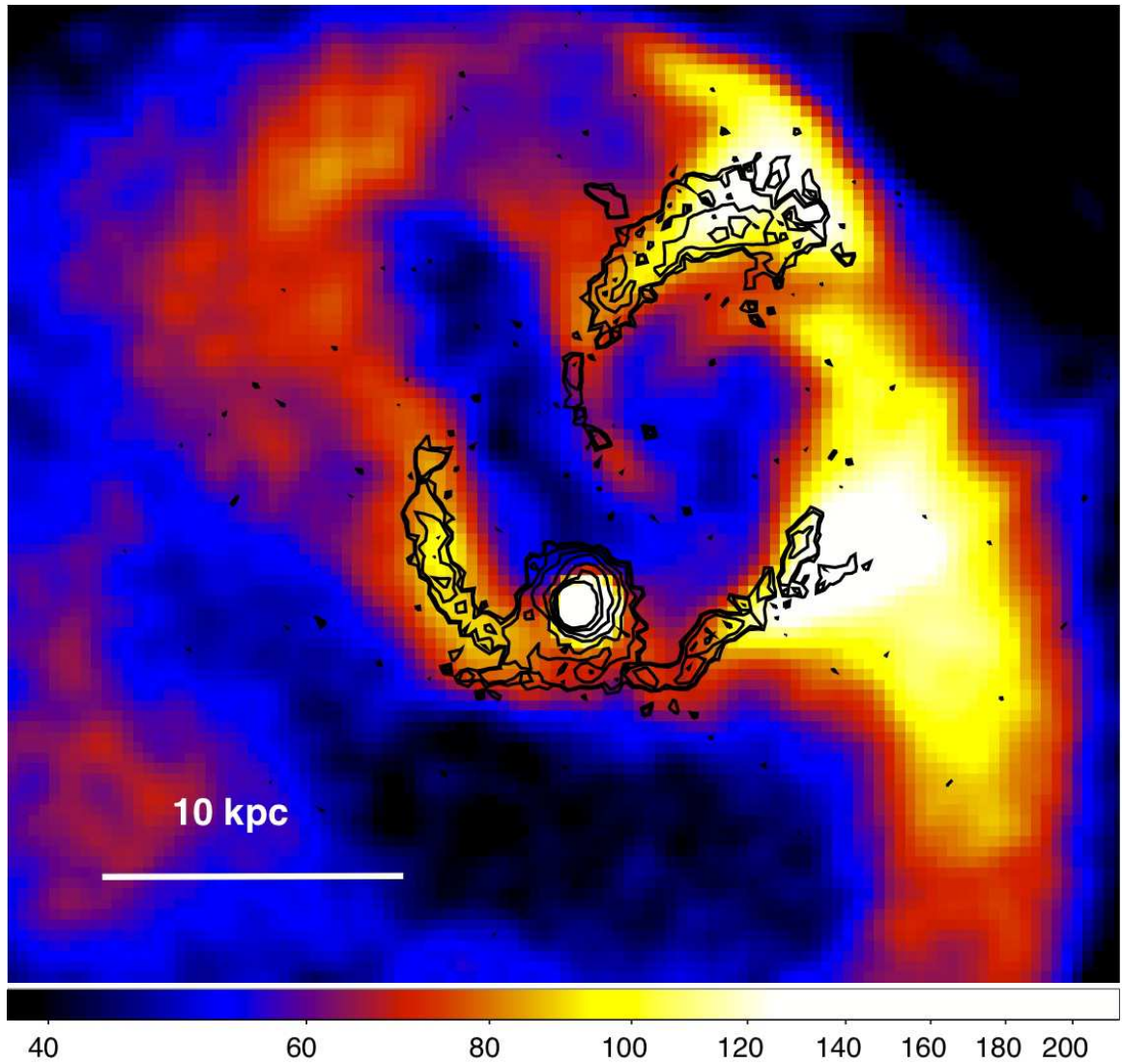}
\caption{{\it Chandra} image in the $0.3-10.0$ keV band, smoothed with a $1\farcs5$ radius Gaussian, and superposed with contours of
$H\alpha$ emission (McDonald et al.\ 2010).  The X-ray brightest regions in the cluster center show excellent correspondence with the
$H\alpha$ emission.
\label{fig:halphaimg}}
\end{figure}

In Fig.\ \ref{fig:Thighrescen}, we display a high-resolution ($0\farcs492$ pixels) temperature map of the central region of A2052, where the 
minimum number of net counts was 2000.  Superposed on the temperature map are X-ray surface brightness contours derived from
the $1\farcs5$ Gaussian-smoothed $0.3-10.0$ keV image.  The coolest parts of the cluster are found in the brightest parts
of the X-ray rims surrounding the bubbles, including the bright rim to the W and NW, the E-W bar that passes through the 
cluster center and AGN, and the N filament that extends into the N bubble.
An overlay of $H\alpha$ contours from McDonald et al.\ (2010)
onto the high-resolution temperature map is shown in Fig.\ \ref{fig:halphatmap}.  For comparison, the $H\alpha$ contours are superposed
onto the $0.3 - 10.0$ keV, $1\farcs5$ radius Gaussian smoothed, {\it Chandra} image in Fig.\ \ref{fig:halphaimg}.  The correspondence between
the $H\alpha$ emission and both the surface brightness and temperature structure is excellent.  
$H\alpha$ emission, representing gas
with $T\approx10^4$ K, is detected in the
brightest and coolest structures in the cluster center, including the E-W bar and the filament that extends into the N bubble.  
While correspondence between $H\alpha$ and X-ray surface brightness was shown in Blanton et al.\ (2001), with the deeper {\it Chandra}
and new $H\alpha$ data presented here, we see more detail in the association, including the filament extending into the N bubble not
reaching the AGN (whereas it appeared to reach the AGN in $H\alpha$ in the Baum et al. (1988) data shown in Blanton et al.\ (2001)).
The association between the $H\alpha$ and X-ray indicates that at least some gas is cooling from the temperature associated with the X-ray gas in these regions ($T\approx10^7$ K) to 
$T\approx10^4$ K.  McDonald et al.\ (2010) and McDonald et al.\ (2011) suggested that this gas represents gas that was previously cooler and then ionized to
produce the $H\alpha$ emission.  Given the low UV-to-$H\alpha$ ratio in this system, they conclude that the ionization source
would likely be shocks related to the AGN rather than nearby, luminous, stars, as in some other cluster centers.

Additionally,
significant clumpy substructure in the temperature distribution is seen in the gas throughout the central cluster region.  This substructure
will be further investigated in an upcoming paper.  Errors
range from approximately $2\%$ in the very central, brightest regions, to approximately $12\%$ in the outer areas of the
map.

\begin{figure}
\plotone{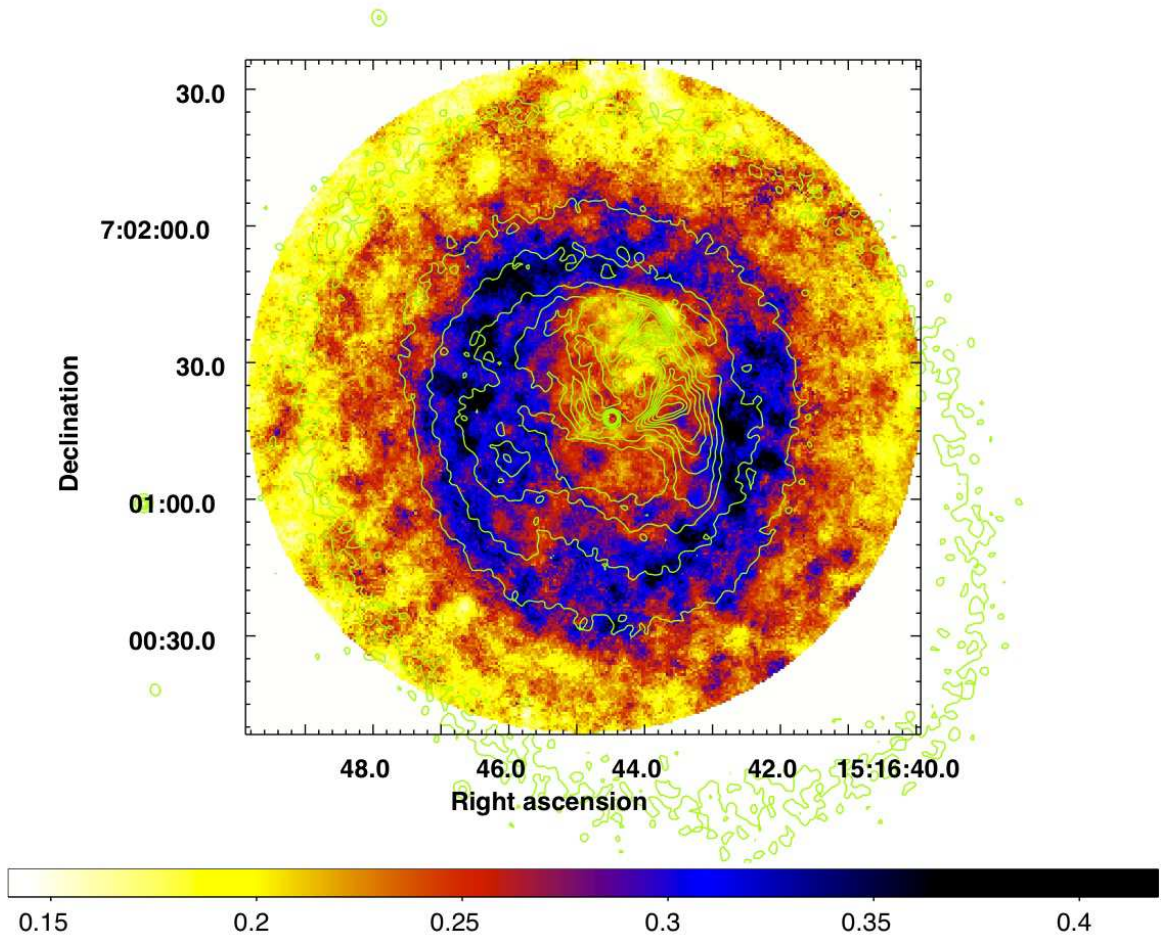}
\caption{Pseudo-pressure map of the central region of A2052 with X-ray surface brightness contours in the $0.3-10.0$ keV range
superposed.  A region of high pressure is seen surrounding the cluster center, outside of the bubble rims, and coincident with
the inner shock feature.
\label{fig:presscen}}
\end{figure}

In Fig.\ \ref{fig:presscen}, we show a projected or ``pseudo'' pressure map of the central region of A2052.  The map was
derived using the APEC normalization and temperature from the spectral fits that resulted in the temperature map in
Fig.\ \ref{fig:Thighrescen}.  The APEC normalization is proportional to $n^{2}V$, where $n$ is density and $V$ is the 
volume projected along the line of sight.  We define the projected pressure as $kT(A)^{1/2}$, where $A$ is the APEC normalization
scaled by area to account for extraction regions that may go off the edge of the chips.   Superposed on the
pressure map are X-ray surface brightness contours in the $0.3-10.0$ keV band.  The most obvious feature in the map is the
clear ring of high pressure that is coincident with the inner discontinuity seen in surface brightness and the jump in 
density (Blanton et al.\ 2009) visible in Figs.\ \ref{fig:unsmoo} and \ref{fig:3color}.  The bubbles, including the inner bubbles to the N and
S of the AGN as well as the outer, small NW bubble and the outer SE bubble, are visible as lower-pressure regions in this projected
pressure map since they are largely devoid of X-ray-emitting gas.

\begin{figure}
\plotone{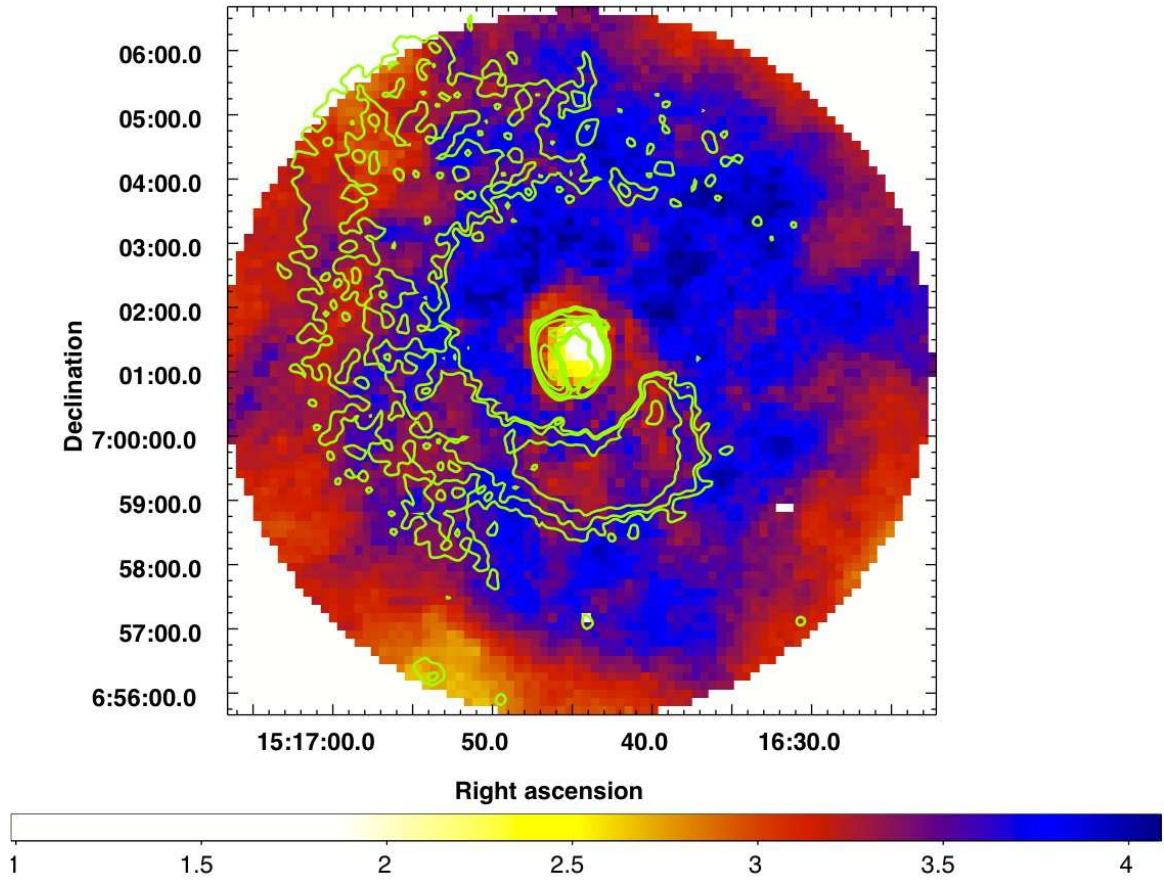}
\caption{Temperature map using a minimum of 10000 background-subtracted counts for each spectral fit.  Contours of residual
surface brightness showing the spiral feature are superposed.  The spiral traces out a region of cooler temperatures.  The
scale bar is kT in units of keV.  Errors range from $1\%$ in the inner regions to $5\%$ in the outskirts of the map.
Holes to the far S and SW are excluded point source regions.
\label{fig:Tmap10000}}
\end{figure}

A lower-resolution temperature map covering a larger region of the center is shown in Fig.\ \ref{fig:Tmap10000}.  
Here, the minimum number
of background-subtracted counts is 10000, and the pixel size is $8\arcsec$.  The errors in temperature range from $1\%$ in the inner
regions to $5\%$ in the outskirts of the frame.  Superposed are the residual surface brightness
contours in the $0.3-2.0$ keV band after beta-model subtraction (see Fig.\ \ref{fig:beta2d}).  The spiral excess traces out a region of temperature lower than its 
surroundings.  In addition, this low temperature feature continues inward along the direction of the spiral toward the cluster
center, beyond the extent of the inner spiral contours.

\begin{figure}
\plotone{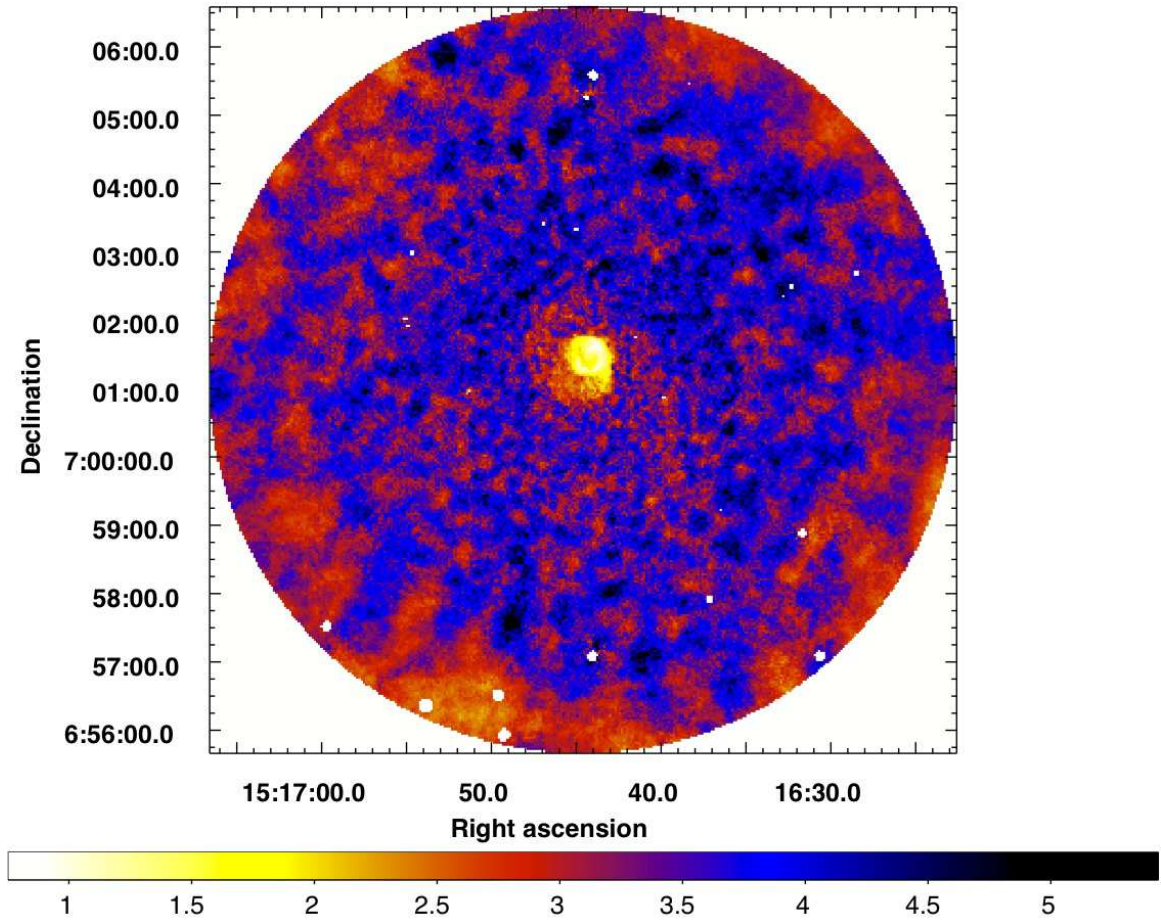}
\caption{Temperature map with the same f.o.v. as Fig.\ \ref{fig:Tmap10000}, but at higher resolution and using a minimum of 2000 
background-subtracted counts for each spectral fit.  The map contains results from more than 80000 spectral fits.
The scale bar is kT in units of keV.  Errors range from $2\%$ in the cluster center to $14\%$ in the outer parts of the frame.
\label{fig:tmap}}
\end{figure}

A higher-resolution temperature map showing the same f.o.v. as in Fig.\ \ref{fig:Tmap10000} is displayed in Fig.\ \ref{fig:tmap}.  The spectra
were extracted with a minimum of 2000 background-subtracted counts, and the pixel size is $4\arcsec$.  The map represents more 
than 80000 spectral fits.  Here, the errors range from $2\%$ in the inner regions to approximately $14\%$ in the outskirts.
Two views of a pseudo-pressure map derived from the fits that were used to make the temperature map in Fig.\ \ref{fig:tmap}
are shown is Figs.\ \ref{fig:pressuremap} and \ref{fig:presszoom}.  The pressure jump corresponding to the innermost shock is
clearly seen, as well as evidence for pressure structure tracing the second inner shock (seen in the outer X-ray contour).
There is no evidence of structure in the pressure map related to the spiral structure.

\begin{figure}
\plotone{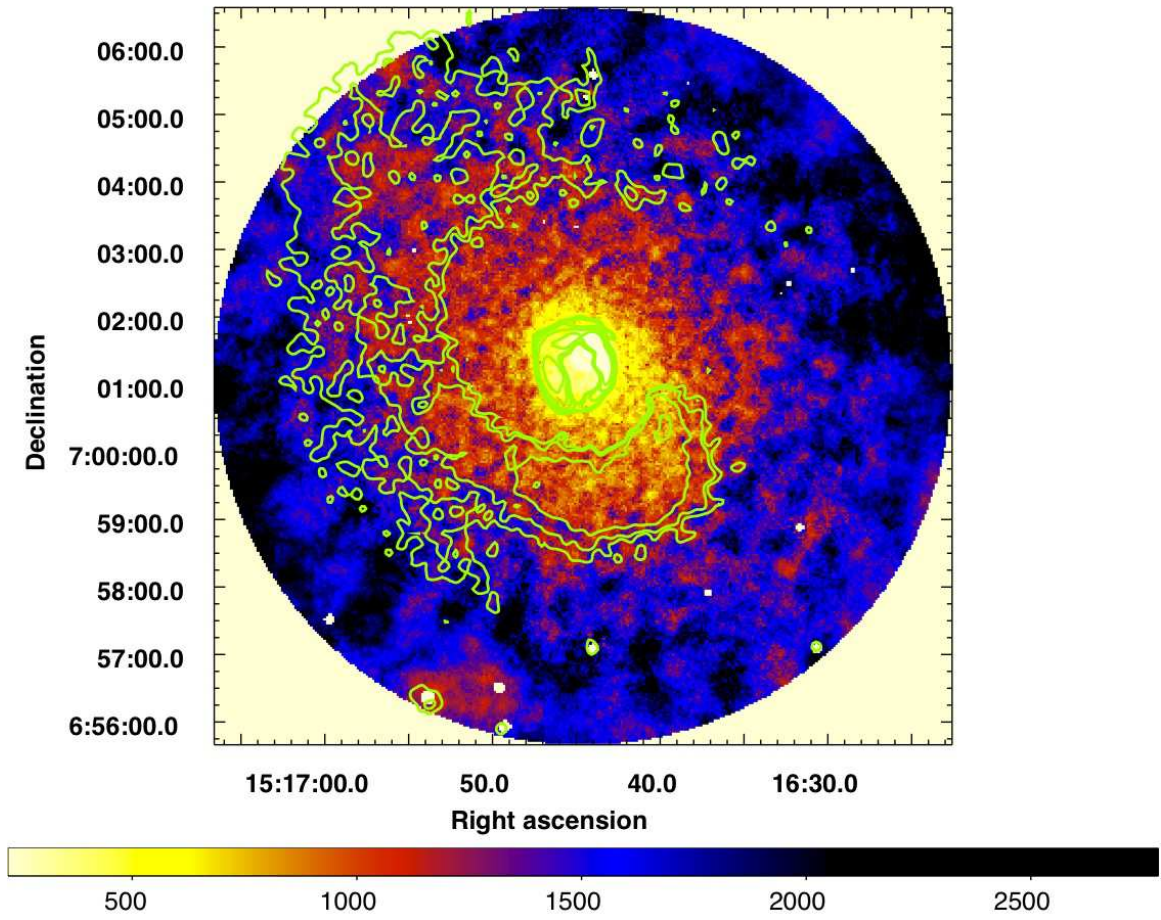}
\caption{Pseudo-entropy map with excess surface brightness contours (after 2D beta model subtraction) superposed.  There is
an overall decrease in entropy towards the cluster center and the sprial feature is conicident with low entropy structure in
the map.
\label{fig:entropymap}}
\end{figure}

A pseudo-entropy map is displayed in Fig.\ \ref{fig:entropymap} with excess surface brightness contours after subtracting a
beta model superposed.  The map was derived using the spectral fits that resulted in the temperature map in Fig.\ \ref{fig:tmap}.
We define pseudo-entropy as $kT(A)^{-1/3}$, with $A$ defined as above.  In general, the entropy decreases toward the cluster center.  
There appears to be correspondence between structure in the entropy map and the spiral feature.

\begin{figure}
\plotone{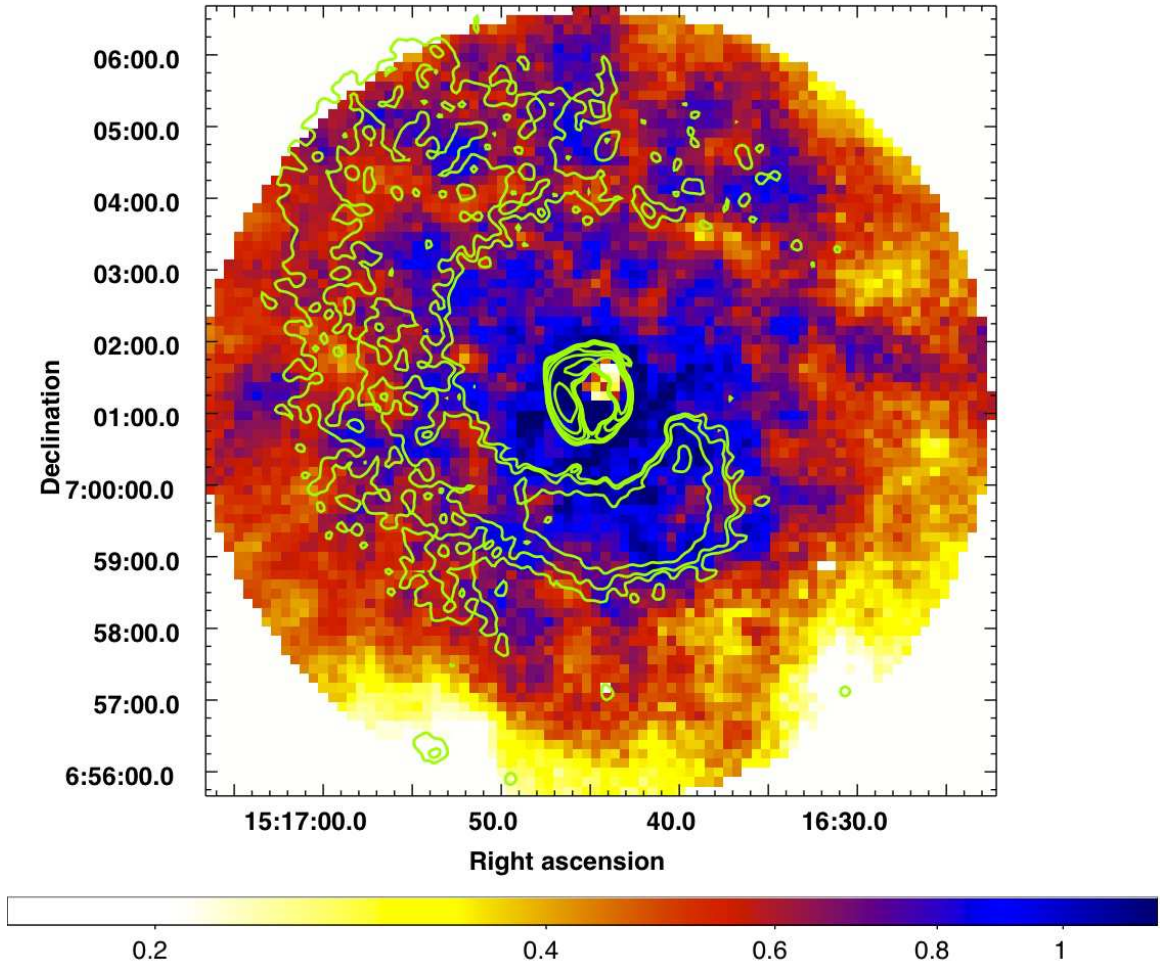}
\caption{Projected abundance map with contours of the spiral excess superposed.  A region of high metallicity is coincident
with the SW portion of the spiral.  Abundances shown are relative to solar.  The region of apparent high abundance 
in the NW 
bubble rim is at least partly the result of fitting a one-temperature model to the projected multi-temperature
gas in this area.  Errors range from $5\%$ in the inner regions to $23\%$ in the outskirts of the frame, with the majority
of the errors at the $10-15\%$ level across the map.
\label{fig:abundmap}}
\end{figure}

A projected abundance map is shown in Fig.\ \ref{fig:abundmap}, corresponding to the temperature map in Fig.\ \ref{fig:Tmap10000}
with $8\arcsec$ pixels and a minimum of 10000 background-subtracted counts per pixel.  Errors range from $5\%$ in the highest surface
brightness regions, to $23\%$ in the outer regions of the map (with the majority of the errors at the $10-15\%$ level across the map).
To the SW, a region of high abundance
is coincident with the high surface-brightness spiral.  This is consistent with higher metallicity gas sloshing away from the
cluster center, creating the spiral.  A high-metallicity region was found at a similar position to the SW using {\it XMM-Newton}
data (de Plaa et a.\ 2010).

\section{Inner Shock Features} \label{sec:shocks}
\begin{figure}
\plotone{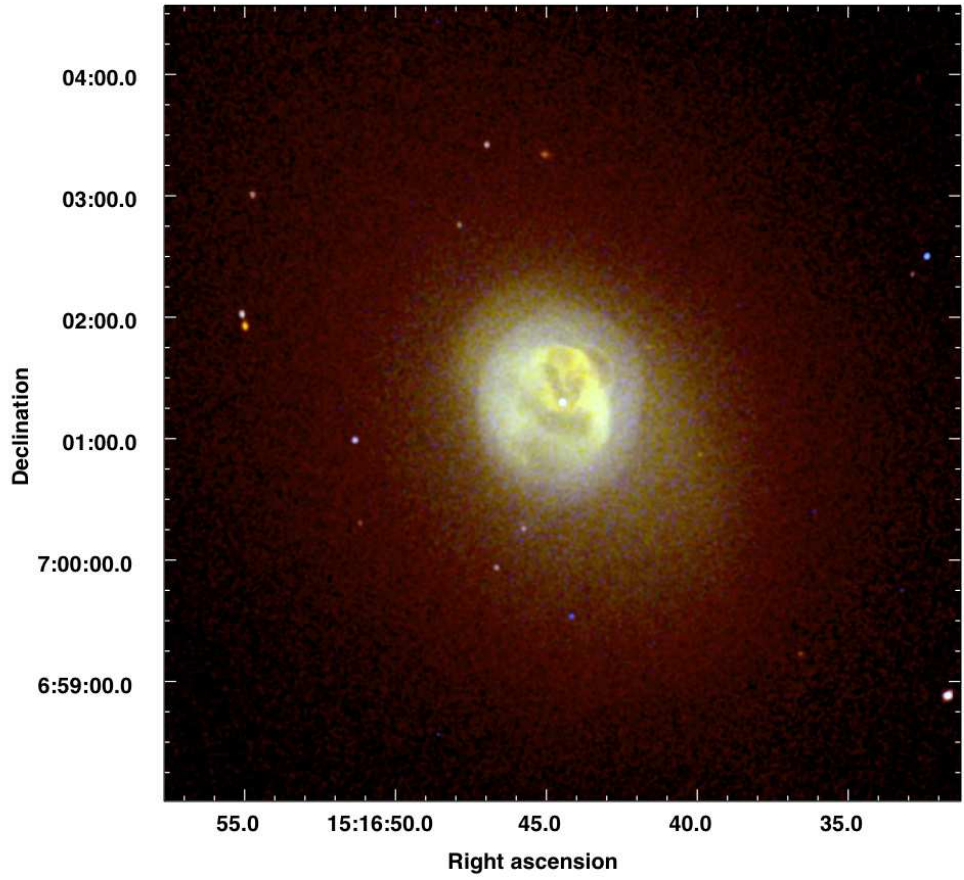}
\caption{Three-color {\it Chandra} image of A2052.  Red is $0.3 - 1.0$ keV, green is $1.0 - 3.0$ keV, and
blue is $3.0 - 10.0$ keV.  Cavities are visible to the north and south of the AGN, surrounded by bright rims.
Exterior to the bright rims, a slightly N-S elliptical shock is seen with hard (blue) emission.
\label{fig:3color}}
\end{figure}

In Blanton et al.\ (2009), we identified two inner surface brightness jumps that were likely associated with shocks.  These jumps are
visible in Figs.\ \ref{fig:3color}, \ref{fig:unsharp}, and \ref{fig:presszoom}.  The first inner jump extends around the cluster center in a
slightly elliptical shape in the N-S direction at a radius of approximately $40\arcsec$.  The second jump is elliptical in the NE to SW
direction, and is sharper and at a smaller radius from the AGN in the NE direction.  To the NE, the second jump is at a radius of approximately
$65\arcsec$.

Similar to  Blanton et al.\ (2009), we have fitted a projected spherical density model to the surface brightness in a NE wedge with PA $-2^{\circ}$ to $98^{\circ}$ measured east from north.
The projected model characterizes the density using power laws and discontinuous jumps.  We use three power laws, and identify two
density jumps.  See Randall et al.\ (2008) for further description of this technique.

Our results are generally consistent with, and a refinement of, those presented in Blanton et al.\ (2009).  
The density jumps are at radii of $44.2\pm{0.1}$ arcsec ($31.2\pm{0.1}$ kpc) and 
$66.3^{+0.3}_{-0.06}$ arcsec ($46.8^{+0.02}_{-0.04}$ kpc), respectively, from the AGN.
The magnitudes of the jumps are factors of $1.25^{+0.016}_{-0.015}$ and $1.29^{+0.010}_{-0.012}$, for the first and second jumps, respectively.
The slopes of the three power-law components, going from the inner to outer regions, are $-0.53^{+0.029}_{-0.024}$, 
$-1.25^{+0.030}_{-0.037}$, and $-1.06^{+0.0067}_{-0.0084}$. 

The jumps in density correspond to Mach numbers of $1.17^{+0.011}_{-0.010}$ and $1.20^{+0.0072}_{-0.0079}$, respectively, for the first
and second jumps.
The temperature is then expected to rise a similar amount inside both shocks (a factor of 1.16 for the first shock and 1.19 for the second shock).

\begin{figure}
\plotone{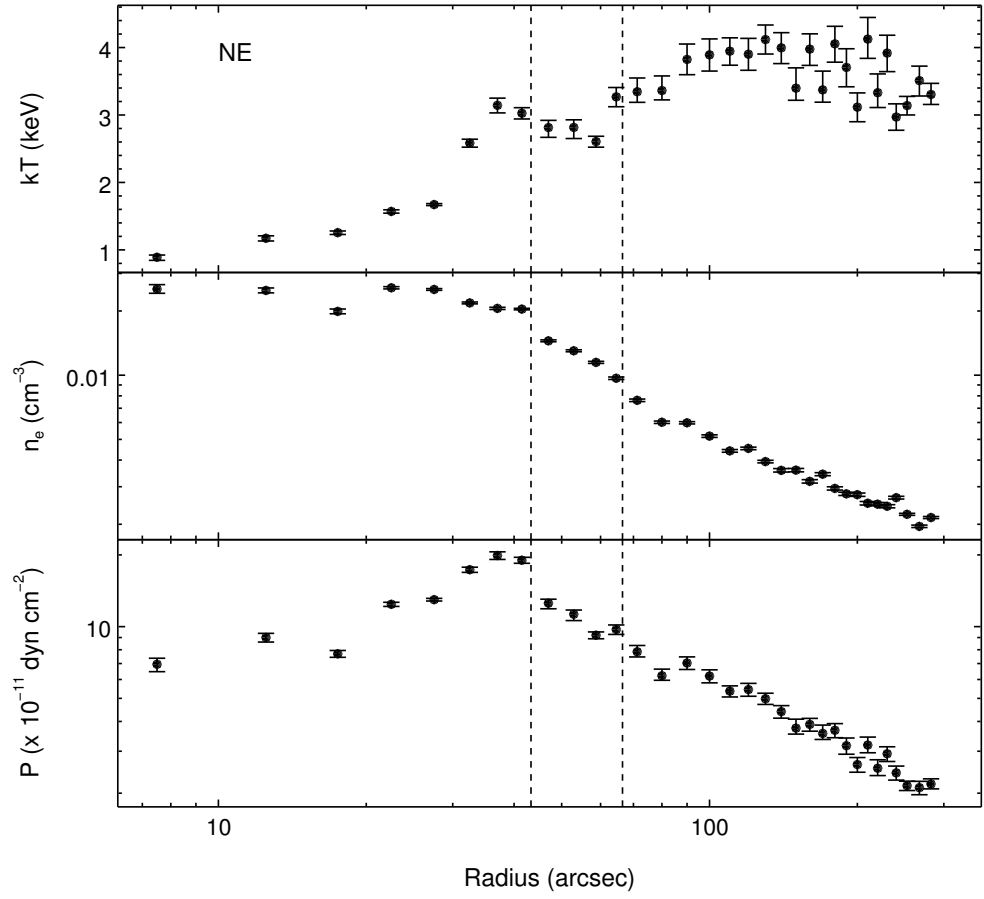}
\caption{Deprojected temperature profile, and profiles of density and pressure for the NE sector.  Positions of the inner shocks are
marked with dashed lines.  A clear rise in temperature and pressure is seen inside the innermost shock.
\label{fig:NE_Tnp}}
\end{figure}

We have calculated a deprojected temperature profile for the NE region, shown in Fig.\ \ref{fig:NE_Tnp}.  Dashed lines indicate the locations
of the two shocks.
As with all of our spectral fits,
spectra were extracted separately for each data set and fitted simultaneously in XSPEC v12.6.  The fits were performed in the $0.6 - 7.0$ keV
range.  A single APEC model plus Galactic absorption was fitted to the outermost annulus spectra with abundance allowed to vary.  
The contribution of emission from this
shell to the next annulus in was calculated using geometric projection, assuming spherically symmetric shells, and the APEC normalization scaled appropriately to account for the
projection of this outer component.  The parameters for the contribution from the outer annulus were frozen, and an additional APEC model
was added for the annulus of interest.  This procedure was continued inward, where the spectral model for each annulus included contributions from all external
annuli.  See Blanton et al.\ (2003) for further description.

In addition, in Fig.\ \ref{fig:NE_Tnp}, we present density and pressure profiles for the NE region.  The density and pressure were determined
from the deprojected spectral fits, since the normalization of the free APEC component for each annulus is proportional to the square of
density at that annulus.
Note that the radius range extends to approximately $300\arcsec$ (212 kpc) where the overdensity (the density relative to the critical density) is
$\approx12500$.  For comparison, an overdensity of 500 ($r_{500}$) is at $r\approx800$ kpc ($\approx1100\arcsec$).

A clear rise in temperature is seen inside the innermost shock, in addition to the jumps in density and pressure in this region.  In
Blanton et al.\ (2009), while the temperatures inside and outside this shock were consistent with the rise expected given the Mach number,
the best-fitting temperatures were approximately flat across the shock.
Here, with the much deeper data set, the rise is clearly detected.  Just outside the shock, the temperature is $kT = 2.81^{+0.11}_{-0.15}$ keV, 
while inside the shock it reaches $kT = 3.14^{+0.11}_{-0.11}$ keV, a factor of $1.12^{+0.10}_{-0.08}$ higher, with a significance
or $2.1\sigma$.  Such temperature rises associated with weak shocks are extremely difficult to measure in cluster centers, with
very deep, high resolution observations required.

For the second shock, the situation is less obvious.  The best-fitting temperatures are appproximately flat across the shock:  
$kT = 3.35^{+0.20}_{-0.16}$ keV outside the shock and $kT = 3.27^{+0.14}_{-0.15}$ keV inside the shock.  Therefore, even within the errors,
this is inconsistent with the expected temperature jump of 1.19, with the highest rise permitted giving a factor of 1.07.
However, as noted above, temperature rises associated with weak shocks are very difficult to detect, and due to projection effects, 
measured rises are
expected to be lower than might be expected based on shock strengths alone (e.g. Randall et al.\ 2011).  
In addition, the gas may cool due to adiabatic expansion (McNamara \& Nulsen 2007).
Also, we note that the temperature drops precipitously in the next annulus inward from the annulus just inside the shock boundary.  This may
indicate that the temperature of the gas just inside the shock boundary was also previously much lower before being
shocked, and then the rise in temperature in this region may be higher than we have estimated above.

In addition, we fitted a projected density model to the surface brightness profile in a wedge to the SW with PA 
$250^{\circ}$ to $340^{\circ}$ from N out to a radius of $100\arcsec$
to characterize the inner shocks in this direction.  We find similar results to the SW as we did to the NE for the first 
inner shock.  We find a density jump of a factor of $1.26\pm{0.025}$ at a radius of $48\pm{0.4}$ arcsec 
($34\pm{0.3}$ kpc).  However, as can be seen in the surface brightness distribution in the images (i.e. Fig.\ 
\ref{fig:3color}), the second inner shock edge seems less sharp and extends to larger radii in the SW than in the NE.  We
do not find a density jump corresponding to the second inner shock to the SW, only a change in slope at a radius of 107 
arcsec (75.5 kpc).

\section{Spiral Feature} \label{sec:spiral}

\begin{figure}
\plotone{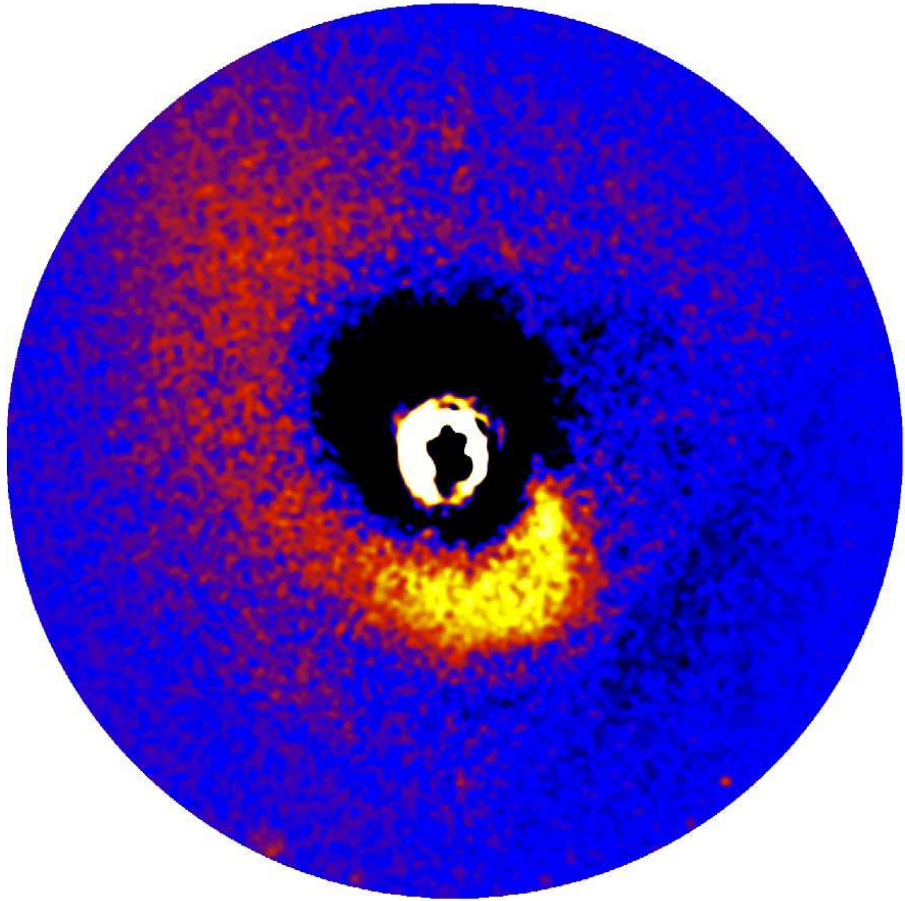}
\caption{Residual image in the $0.3 - 2$ keV band of the central $5\farcm66$ (240 kpc) radius region of A2052 resulting from the subtraction 
of a 2D beta model.  The image
has been smoothed with a $7\farcs38$ Gaussian.  In addition to the bubble rims seen in the center of the image, on larger 
scales, a spiral is visible extending from the SW to the NE.  The linear feature in the SW is a chip-edge artifact.
\label{fig:beta2d}}
\end{figure}

\begin{figure}
\plotone{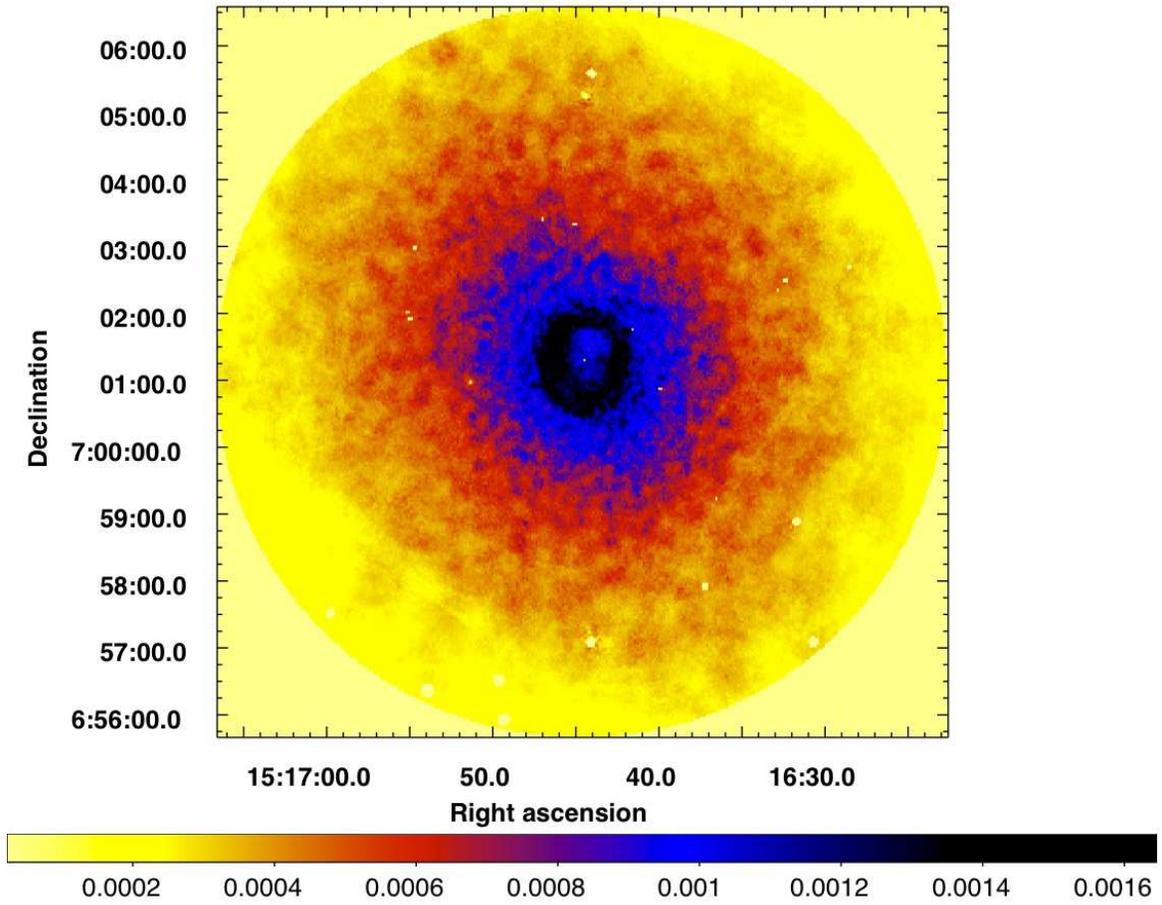}
\caption{Pseudo-pressure map using the spectral fits that resulted in the temperature map in Fig.\ \ref{fig:tmap}.  As in
Fig.\ \ref{fig:presscen}, the slightly N-S elliptical region corresponding to the inner shock is seen.  There is no evidence
for a region of pressure that corresponds with the spiral feature seen in Fig.\ \ref{fig:beta2d}.
\label{fig:pressuremap}}
\end{figure}

\begin{figure}
\plotone{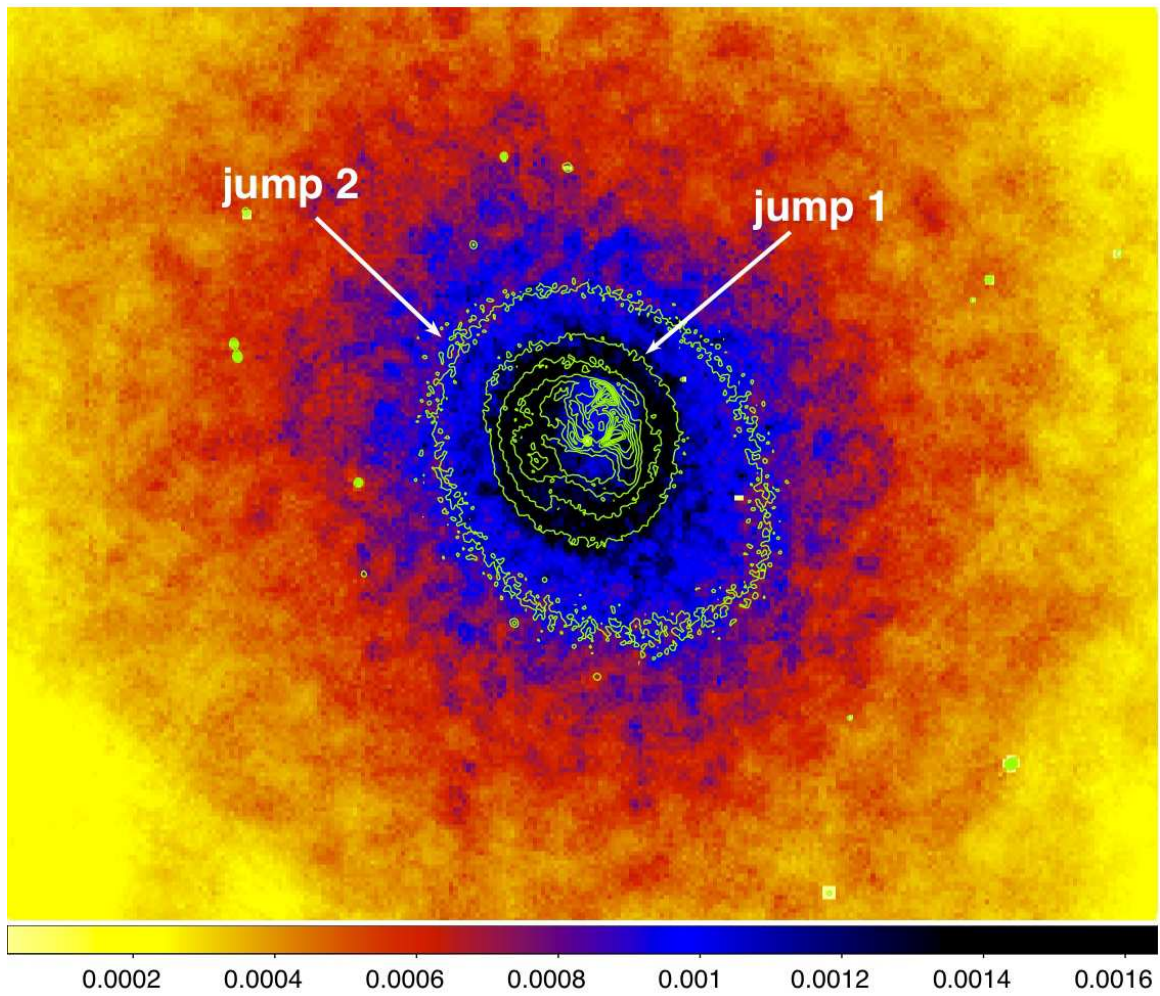}
\caption{Zoomed-in view of the pseudo-pressure map shown in Fig.\ \ref{fig:pressuremap} with X-ray surface brightness contours
superposed.  The first inner jump in surface brightness traces out a jump in pressure, and the second inner surface
brightness jump traces out a region of enhanced pressure.  
\label{fig:presszoom}}
\end{figure}

The spiral feature visible in Fig.\ \ref{fig:beta2d} is similar to that seen in simulations of gas sloshing
in cluster centers (Ascasibar \& Markevitch 2006).  The sloshing sets up cold fronts and the distinctive spiral morphology.
We therefore would expect the excesses seen in Fig.\ \ref{fig:beta2d} to be visible as excesses in surface brightness profiles 
containing these regions.  Temperature profiles should show cooler gas coincident with the surface brightness 
enhancements.  In addition, the simulations predict that the bright, sloshing, spiral region will contain gas of lower entropy as cluster
central gas is displaced to larger radii.  As shown in \S\ref{sec:maps}, the spiral excess is coincident with regions of low temperature and entropy in 
projected maps.

\begin{figure}
\plotone{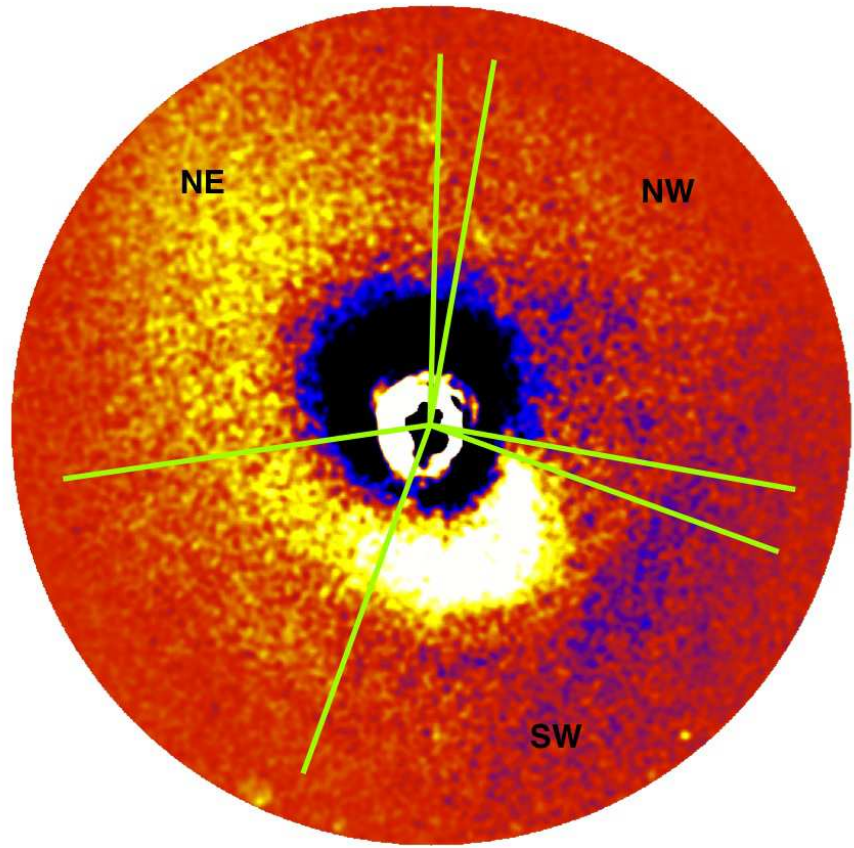}
\caption{Residual image in the $0.3 - 2$ keV band of the central $5\farcm66$ (240 kpc) radius region of A2052 resulting from the subtraction 
of a 2D beta model and smoothed with a $7\farcs38$ Gaussian.  Superposed on the image are the SW, NE, and NW sectors from which
the surface brightness and temperature profiles shown in Figs.\ \ref{fig:NE_Tnp}, \ref{fig:spiralsurfbr}, \ref{fig:kTSWNW}, 
\ref{fig:kTNENW}, and \ref{fig:SWdepro} were extracted.
\label{fig:beta2dreg}}
\end{figure}

\begin{figure}
\includegraphics[angle=-90,scale=0.47]{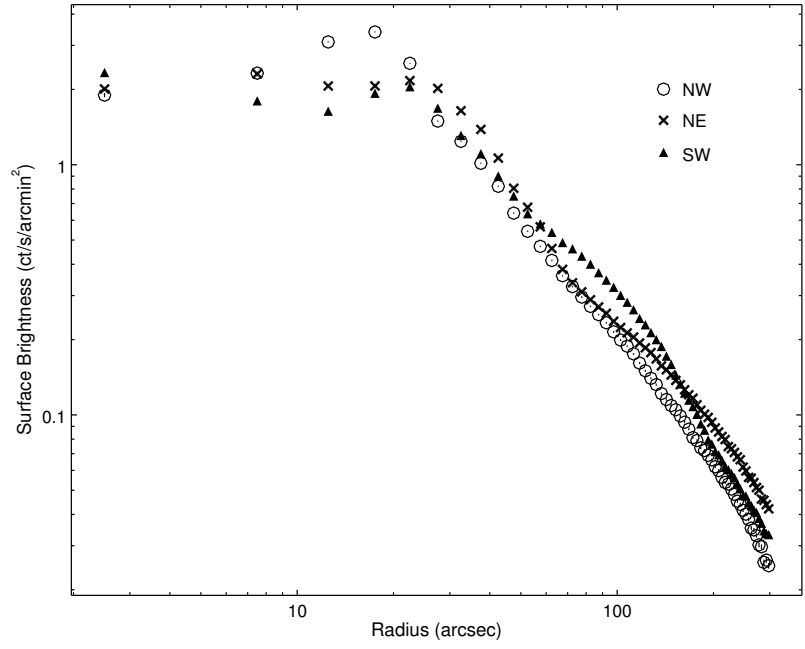}
\caption{Surface brightness profiles in the $0.3 - 10$ keV band corresponding to the sectors shown in Fig.\ \ref{fig:beta2dreg}.
Clear excesses are seen in the regions of surface brightness enhancement associated with the spiral feature in Figs.\
\ref{fig:beta2d} and \ref{fig:beta2dreg}.  For the SW, the excess extends from $\approx60\arcsec - 195\arcsec$, while for
the NE, the excess is seen at radii beyond $\approx110\arcsec$.
\label{fig:spiralsurfbr}}
\end{figure}

We have extracted surface brightness profiles from wedges in three directions:  SW, NE, and NW.  The SW corresponds to the bright,
inner part of the spiral, while the NE includes the outer region of the spiral, and the NW is largely free from any
spiral excess emission and serves as a comparison region.
The sectors used for the profiles are shown superposed on the $0.3 - 2.0$ keV residual image in Fig.\ \ref{fig:beta2dreg}.  The surface 
brightness profiles for the three regions are shown in Fig.\ \ref{fig:spiralsurfbr}.  Error bars are smaller than the 
symbols.
Relative to the NW, non-spiral region, the SW shows excess emission from approximately $60\arcsec - 195\arcsec$
(42 -- 138 kpc), corresponding
to the inner part of the sprial.  To the NE, excess emission is seen beyond approximately $110\arcsec$ (78 kpc).  Note that the enhancement
in the NW in the $\approx10 - 20\arcsec$ (7 -- 14 kpc) region shows that the bubble rims are brightest in this region, which also corresponds
to cooler gas seen in $H\alpha$ (Figs.\ \ref{fig:halphatmap}, \ref{fig:halphaimg}).

\begin{figure}
\includegraphics[angle=-90,scale=0.8]{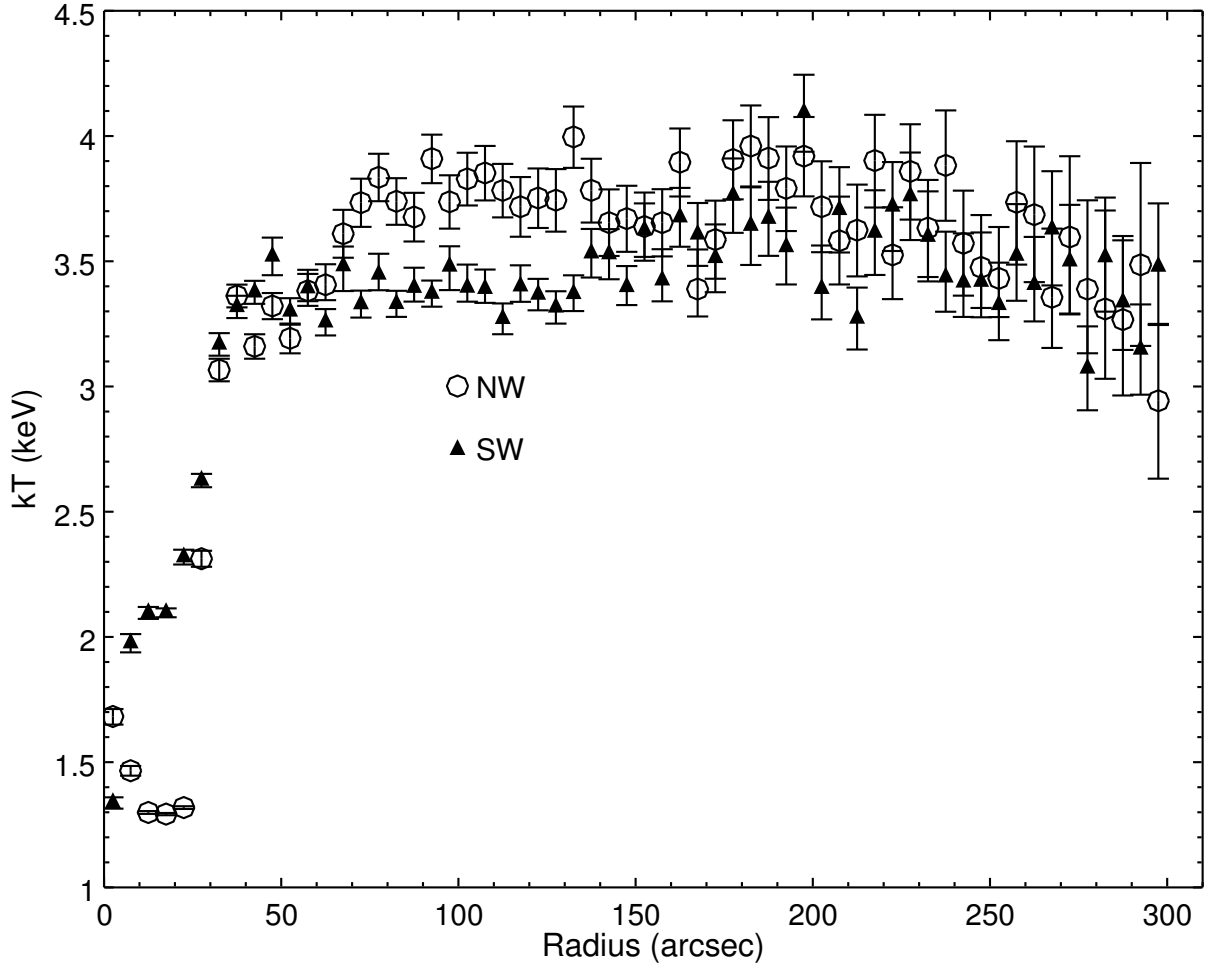}
\caption{Projected temperature profiles for the SW and NW sectors shown in Fig.\ \ref{fig:beta2dreg}.  The SW sector corresponds with
the bright, inner, spiral region, while the NW sector is the non-spiral, comparison region.  The SW spiral excess 
corresponds with regions of cooler temperature in the radial range $\approx60\arcsec - 140\arcsec$.  The profiles converge beyond
$\approx140\arcsec$.
\label{fig:kTSWNW}}
\end{figure}

\begin{figure}
\includegraphics[angle=-90,scale=0.8]{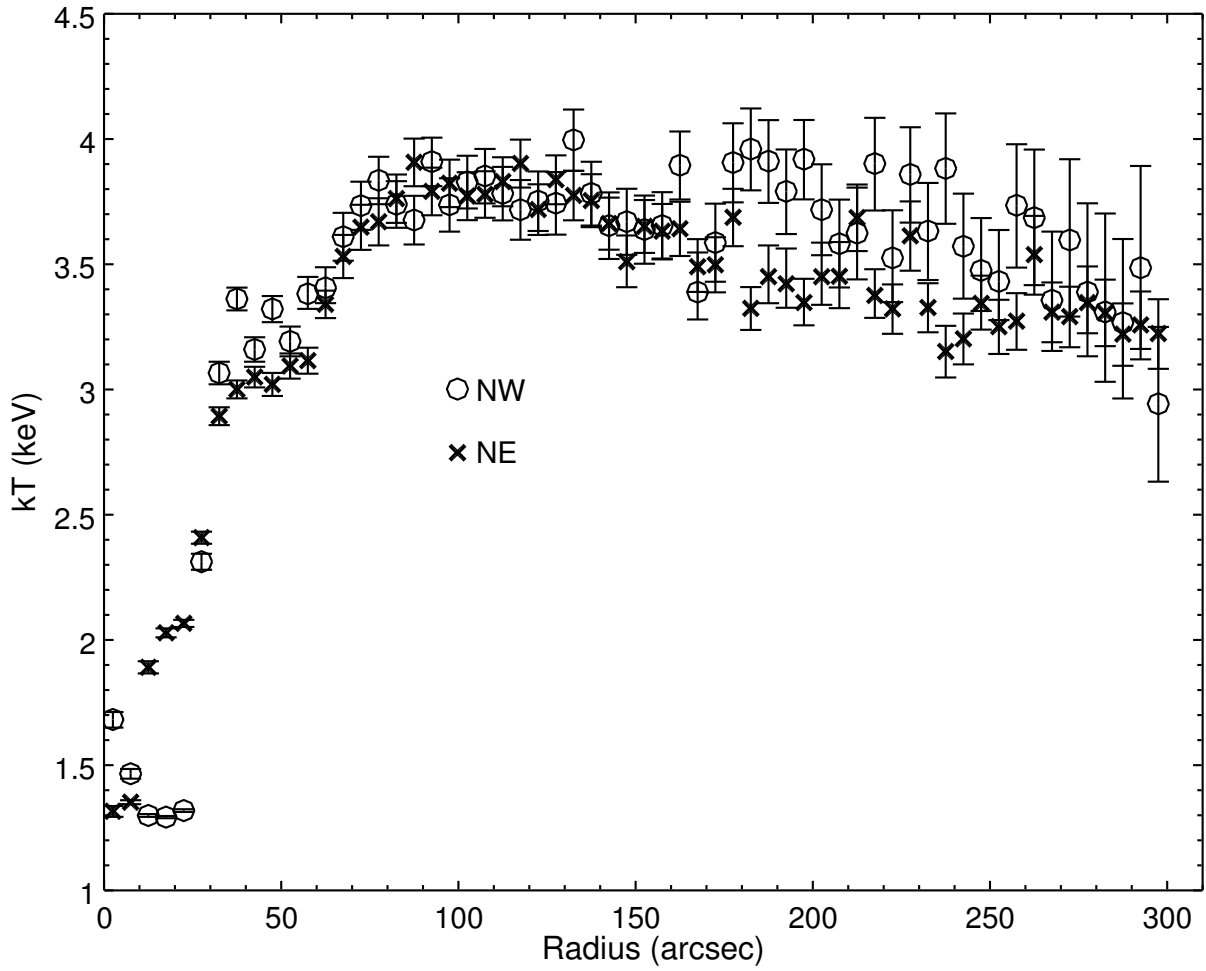}
\caption{Projected temperature profiles for the NE and NW sectors shown in Fig.\ \ref{fig:beta2dreg}.  The NE sector corresponds with
the outer spiral region, while the NW sector is the non-spiral, comparison region.  Cooler temperatures are seen associated
with the NE spiral excess (radii $>160\arcsec$).
\label{fig:kTNENW}}
\end{figure}

We have extracted spectra in the three wedges and determined projected temperature profiles.
The spectra were extracted separately for each of the observations as well as for the corresponding background files.  The spectra for
each region
were fitted simultaneously using an APEC thermal plasma model.
Absorption was fixed at the Galactic value and elemental abundances were allowed to vary.
Projected temperature profiles comparing the SW and NW (non-spiral) regions and the NE and NW regions are shown in Figs.\ \ref{fig:kTSWNW} and
\ref{fig:kTNENW}, respectively.  In both cases, significant drops in temperature are seen in the regions corresponding to
the surface-brightness excesses associated with the spiral described above.
In addition, the high surface brightness $\approx10\arcsec - 20\arcsec$ (7 -- 14 kpc) region in the NW has cooler temperatures than those radii regions
to the SW and NE.  This is likely due, at least in part, to the larger contribution to the emission measure from the bright, compressed, cool, bubble rims to
the NW at these radii as compared to the same radii to the SW and NE, where the bubble rims are not as bright.

In order to examine the bright region of the spiral to the SW in more detail, we have fitted a projected density model
to the surface brightness profile, as described above for the NE sector shock fits.  To focus on the cold front / spiral edge, we
fit only regions with $r > 70\arcsec$.  We find a density jump of a factor of $1.12^{+0.022}_{-0.011}$ at a radius
of $154\pm{1.5}$ arcsec ($109\pm{1}$ kpc) from the cluster center.

\begin{figure}
\plotone{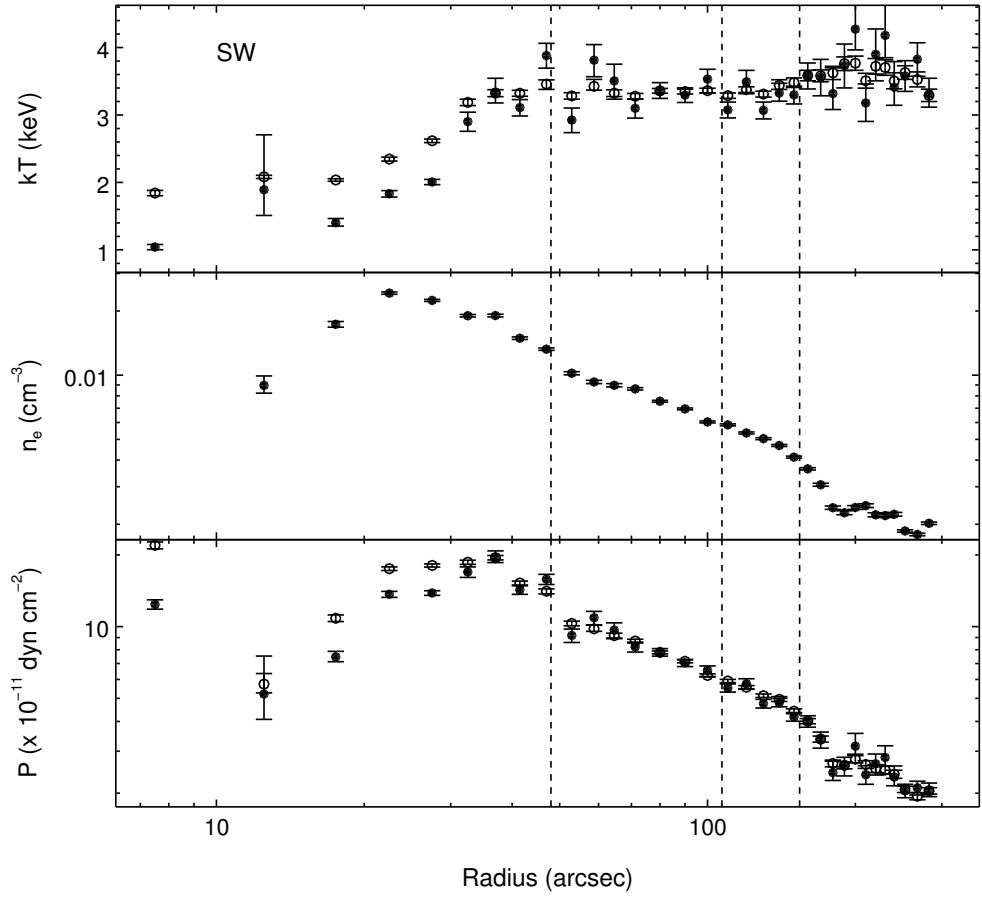}
\caption{Profiles of temperature, density, and pressure in the SW sector.  Filled circles indicate results from deprojected
fits.  The open circles show the projected temperature profile and the pressure profile derived using the density
and the projected temperatures.  Dashed lines show the positions of the first inner shock, the second inner shock / edge,
and the cold front / SW spiral edge, going from smaller to larger radii.
\label{fig:SWdepro}}
\end{figure}

We extracted spectra in larger bins to the SW, and performed a spectral deprojection as described above for the NE.  
Profiles of temperature, density, and pressure are shown in Fig.\ \ref{fig:SWdepro}.  Dashed lines, going from smaller
to larger annuli, indicate the positions of the first inner shock, the second inner shock / edge, and the exterior edge
of the cold front / SW spiral.
Since the deprojection introduces some scatter, particularly in the region between the inner shocks, we have also
plotted the projected temperature profile as well as the pressure profile using the projected temperatures (shown as open
circles).
A temperature rise is seen associated with the first inner shock, but not the second inner shock.  A clear density
fall off is seen at the outer edge of the SW spiral.  The temperatures within the sprial region are cooler than those
outside of it.

\begin{figure}
\plotone{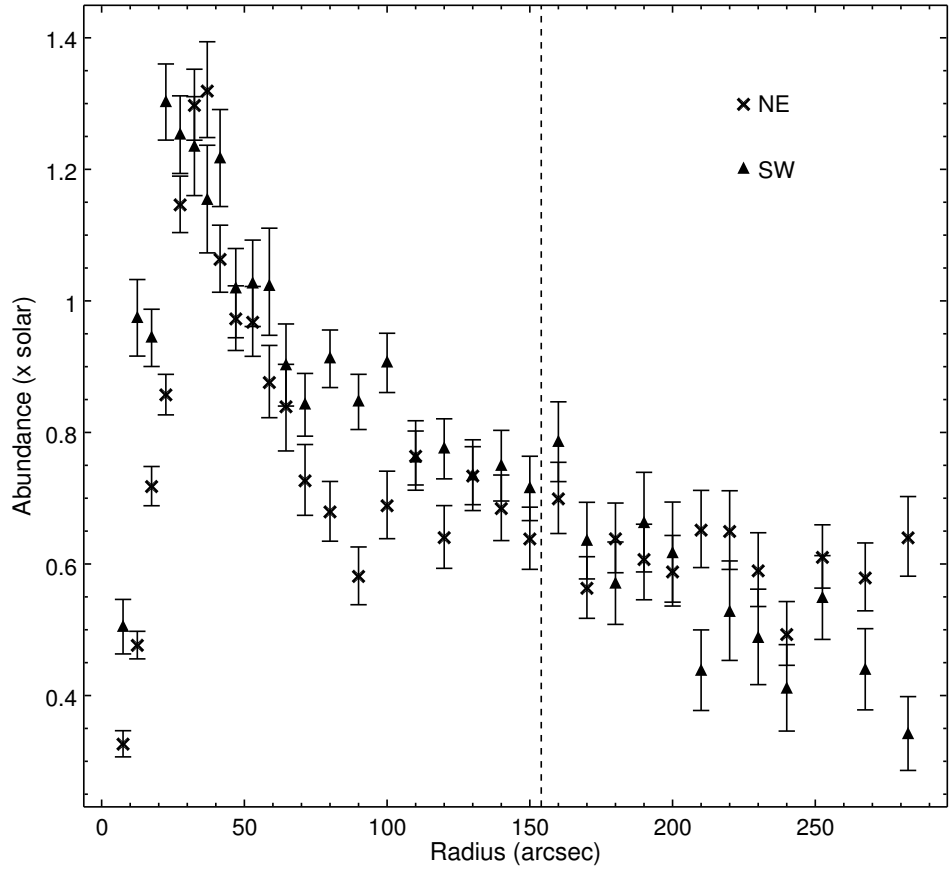}
\caption{Projected abundance profiles for the SW and NE regions.  The dashed line indicates the outer edge of the SW
spiral / cold front.  The abundances are higher inside the SW spiral region (from $\approx70\arcsec - 120\arcsec$) compared to the corresponding radial region to
the NE.
\label{fig:abundprof}}
\end{figure}

\begin{figure}
\includegraphics[angle=-90,scale=0.5]{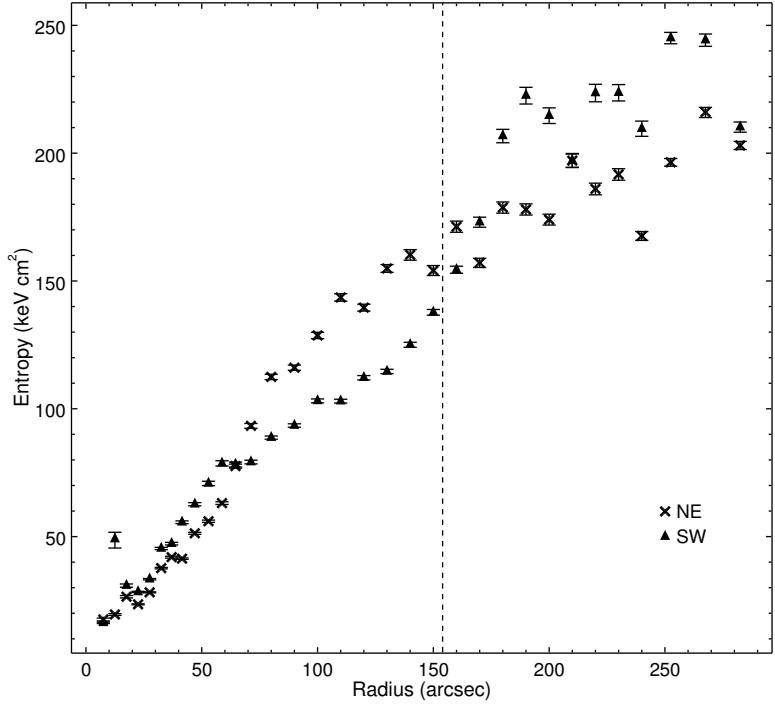}
\caption{Entropy profiles for the SW and NE regions with entropy defined as $S=kT/n_{e}^{2/3}$, and using projected temperatures.  The dashed line indicates the outer 
edge of the SW spiral.  The spiral surface-brigthness excess is very well traced by the entropy structure.
The spiral excess to the SW ($r\approx70-150\arcsec$) corresponds with a region of lower entropy.  At larger radii
$r>190\arcsec$, the spiral excess in the NE also is shown to have lower entropy values as compared to the non-spiral region in the SW at these
large radii.
\label{fig:entropyprof}}
\end{figure}

Projected abundance profiles to the SW and NE are shown in Fig.\ \ref{fig:abundprof}.  The dashed line marks the outer
edge of the spiral to the SW (at $r=154\arcsec$, as determined above).  The abundance is higher in the bright, SW, spiral region, 
than in the corresponding radial region to the NE.  
Profiles of entropy for the SW and NE regions are shown in Fig.\ \ref{fig:entropyprof}.  The entropy is defined as $S=kT/n_{e}^{2/3}$, and
projected temperature values were used.  As in Fig.\ \ref{fig:abundprof}, the outer edge of the SW spiral region is marked with a dashed line.
There is clearly a cross-over in the entropy profiles for the SW and NE regions near this radius.  From $r\approx70-150\arcsec$, low entropy
values are found coincident with the SW spiral excess.  In the NE, the spiral excess is at larger radii, and this is seen in the entropy profile,
where the values are low in the NE with radii greater than approximately $190\arcsec$.  This is striking confirmation of the predictions from sloshing
models, where central, low entropy, cluster gas is displaced to larger radii (i.e.\ Ascasibar \& Markevitch 2006).

The sum of the evidence, including the surface brightness profiles, and temperature, abundance, and entropy distributions, points
to the spiral being a cold front / sloshing feature resulting from an off-axis merger earlier in the cluster's history.

\section{X-ray Cavities}

\begin{figure}
\plotone{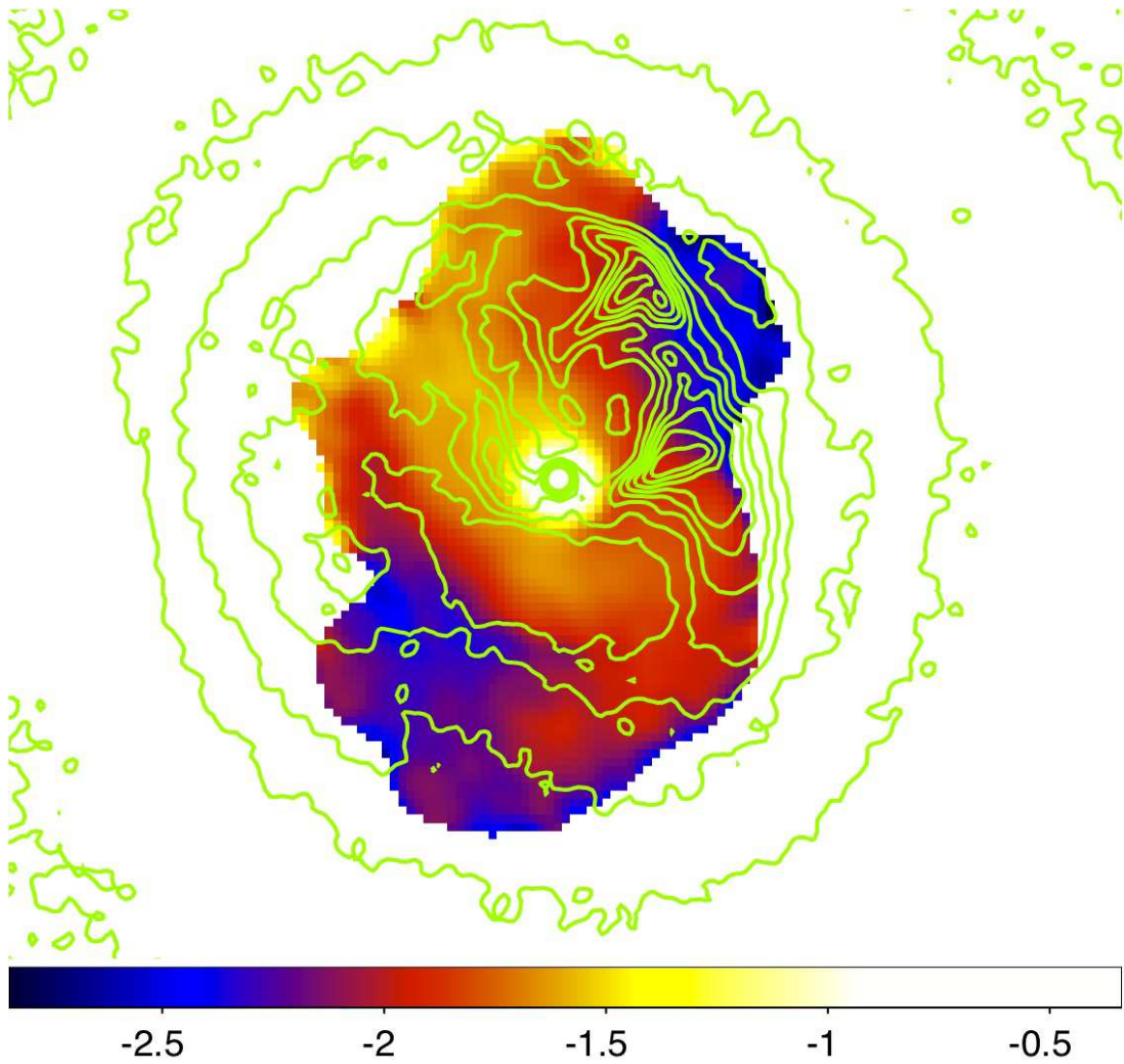}
\caption{Radio spectral index map created using the 1.4 and 4.8 GHz data.  Contours of $0.3 - 10.0$ keV X-ray surface brightness
are superposed.  The outer bubbles to the NW and SE are filled with radio emission with steeper spectral index, consistent with these
regions being inflated by an earlier outburst of the AGN.
\label{fig:specindex}}
\end{figure}

Clear cavities (or ``bubbles'') in the X-ray emission are seen in the {\it Chandra} images to the N and S of the AGN.  In addition, there
is a small bubble bounded by a narrow X-ray filament to the NW and a bubble separated from the S bubble to the SE.
A very small depression is also seen to the E.  These features are all filled with 4.8 GHz radio emission as seen in
Fig.\ \ref{fig:unsharp}.  As described in \S\ref{sec:radio}, we created a radio spectral index map using the 1.4 and
4.8 GHz VLA data.  The map is shown in Fig.\ \ref{fig:specindex} with contours of $1\farcs5$ Gaussian-smoothed X-ray 
emission in the $0.3 - 10.0$ keV band superposed.  

The spectral index is flattest at the position of the radio and X-ray core, and steepens into the radio lobes.  
The regions we
have identified as outer bubbles, to the NW and SE, are associated with regions with distinctly steeper spectral
indices.  This is consistent with these bubbles resulting from an earlier stage in the current outburst, or from a separate, earlier
AGN outburst, since the high-energy electrons age faster than the lower energy electrons, resulting in steepening of the 
spectrum over time.  The radio emission ``leaking'' out of the N bubble through the bubble rim directly N of the AGN has
a spectral index consistent with that filling the N lobe, making it likely that this emission is part of the current
AGN outburst.

\begin{figure}
\plotone{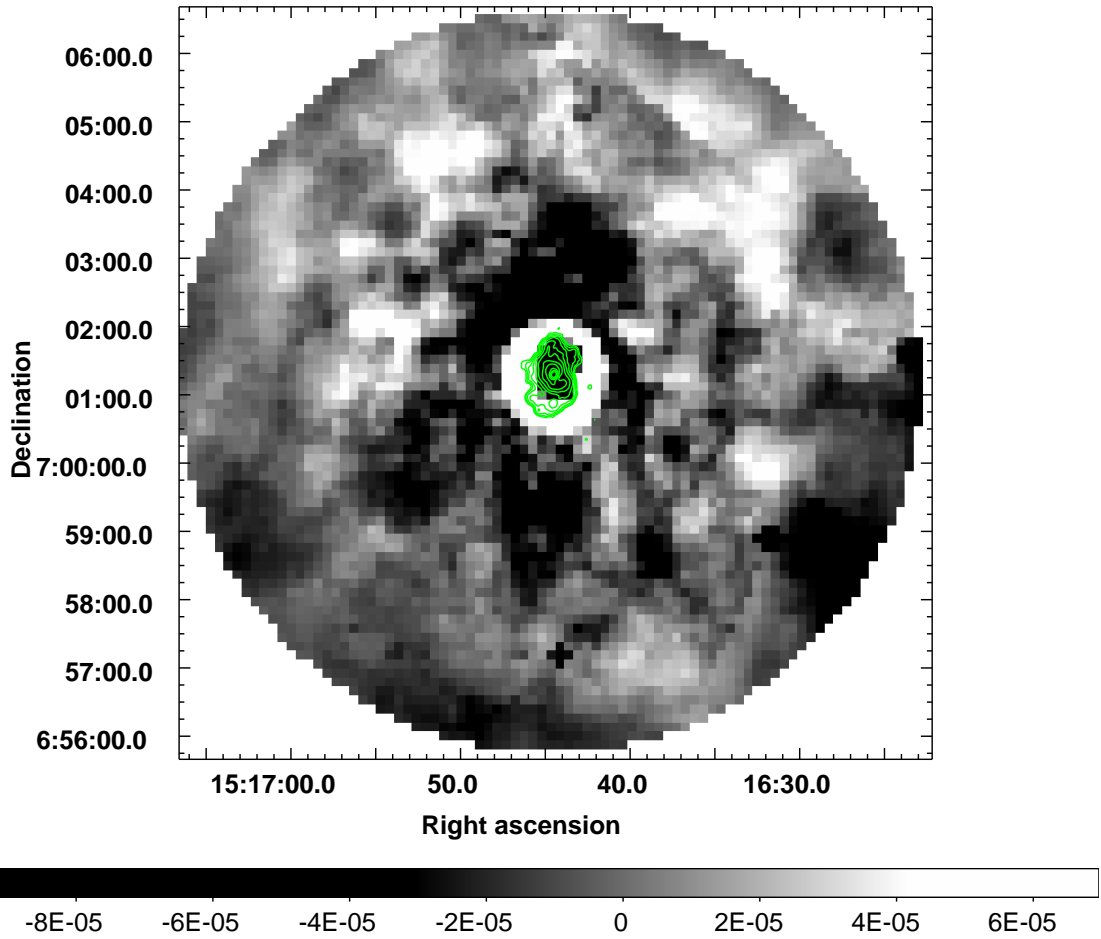}
\caption{Pressure residual map created by subtracting off a 2D beta model fitted to a pseudo-pressure map with 
4.8 GHz radio contours superposed.  The inner shock is seen as a pressure enhancement exterior to the radio lobes.
In addition, large deficits in pressure are seen to the N and S of the cluster center.
These may represent ``ghost cavities'' related to an earlier AGN outburst.
\label{fig:pressdiff}}
\end{figure}

Obvious additional bubbles are not seen outside of this central region in the images, including the unsharp-masked
images.  Also, lower frequency radio emission at 330 MHz (Zhao et al.\ 1993) has a similar extent as the 1.4 and 4.8 GHz 
emission.  
To search for more bubbles at larger radii, we have made a pressure-difference map.
The map was created by fitting a 2D beta model to a pressure map with $8\arcsec$ bins, and a minimum of 10000 
background-subtracted counts for each region used for a spectral fit.  This corresponds to the temperature map shown
in Fig.\ \ref{fig:Tmap10000}.
The center was fixed to the position found in the 2D beta model fit to the surface brightness, described in
\ref{sec:beta}.  The values for the ellipticity and position angle of the ellipse, 0.17 and $119^{\circ}$ from W,
respectively, were similar to those found for the surface brightness fit.
The pressure-difference map is shown in Fig.\ \ref{fig:pressdiff}, with 4.8 GHz radio contours superposed.
Regions of low pressure are evident to the N and S of the AGN, along the current axis of the radio source.
For both the N and S apparently low-pressure regions on the map, we have extracted spectra within circular apertures covering
the regions, as well as comparison regions in surrounding circular annuli.  We have calculated ``projected'' pressures, 
defined as $kT(A)^{1/2}$, where $A$ is the APEC normalization, for the apparent bubbles (low pressure regions) and the comparison
annuli.  For the N bubble, we find a significance in the lower projected pressure of $2.3\sigma$ compared to the comparison annulus,
and for the S bubble, the significance is $2.2\sigma$.
These regions of lower pressure may represent outer bubbles from an earlier outburst (or multiple earlier outbursts) of the AGN.
Future higher dynamic range radio data may give further evidence that these features are related to AGN activity.

Approximating both of these possible bubbles as spheres with radii of 34 kpc, located 75 kpc from the cluster center,
we calculate their energy input into the ICM as $E = 4PV$ (Churazov et al.\ 2002).  This assumes that the bubbles are filled
with relativistic plasma ($\gamma=4/3$).  Using the average pressure at the cluster radius of 75 kpc of $P\approx6\times10^{-11}$
dyn cm$^{-2}$, we find that each bubble can add $1\times10^{60}$ erg to the ICM.

\section{Discussion and Conclusions}

We have presented first results from a very deep {\it Chandra} observation of Abell 2052.  Detailed structure
is seen in the inner part of the cluster, including bubbles, bright shells, and filaments.  Radio emission
at 4.8 GHz is seen to fill bubbles to the N and S of the AGN, as well as outer bubbles to the NW and SE, and a
small bubble to the E.  $H\alpha$ emission is coincident with the brightest and coolest regions in the cluster center,
including the E-W bar and the filament in the N bubble, indicating that at least some gas is cooling to $T\approx10^4$ K.

An inner shock is
clearly seen surrounding the cluster center, as well as a second feature exterior to the first that is also
most likely a shock.  Both edges in surface brightness can be described as arising from shocks with Mach numbers
$\approx1.2$.  For the inner shock, evidence for an associated temperature rise is seen in a three-color image.
Also, a temperature profile reveals a significant increase inside this region, consistent with that expected given the
shock strength.  Such temperature rises associated with weak shocks
driven by AGN in the centers of cool core clusters have only rarely been measured, given the narrow widths of the shocks, 
the small magnitudes of the temperature rises, and the cooling related to
adiabatic expansion and the projection of cluster gas at larger radii.

A spiral feature is seen, and is well-described as resulting from gas sloshing
related to a cluster-cluster or sub-cluster merger earlier in the lifetime of A2052.  
There is also evidence of dynamical activity in the cluster from optical spectroscopy:  the 
central cD has a fairly large peculiar velocity ($290\pm90$ km s$^{-1}$) relative to the cluster mean (Oergerle \& Hill 2001).
The X-ray spiral is brightest
to the SW, and continues around the cluster center to the E and NE.  The SW portion of the excess emission
was shown in Lagan\'{a} et al.\ (2010) using a shorter {\it Chandra} exposure, and in de Plaa et al.\ (2010)
using a deep {\it XMM-Newton} observation.  The spiral structure was not seen, however, before this very deep
{\it Chandra} observation.  Lagan\'{a} et al.\ presented several residual {\it Chandra} images of cool core clusters, and noted that A2052 was an exception in not exhibiting a spiral feature in the available data, and de Plaa
et al.\ described the SW excess as a cold front.
Here, in addition to exhibiting an excess in surface brightness on a 2D 
model-subtracted {\it Chandra} image, the brightness distribution is confirmed using surface brightness profiles in different
sectors across the image.  The spiral is also shown to contain gas cooler than its surroundings in both temperature
maps and sector profiles.  Regions of low entropy are found coincident with the spiral surface-brightness excess.
Finally, the abundance is higher in the spiral region than its surroundings.  
This is consistent with
central, low-entropy, high-metallicity gas sloshing out to larger radii.  Therefore, sloshing can play a part in redistributing
the metals within a cluster (Simionescu et al.\ 2010, de Plaa et al.\ 2010).  Outbursts from the AGN also play a role in metal redistribution (Kirkpatrick et al.\ 2009).  Detailed measurements of the spectral properties of similar diffuse spiral excesses have only rarely been
measured.  While Lagan\'{a} et al.\ (2010), for example, presented evidence of excess surface-brightness spirals in several
systems, in many cases temperature declines were not detected associated with the spiral features, and if they were, they
were most often only measured for the bright, inner spiral regions.

While the radio emission shows clear correspondence with deficits in the X-ray emission, and the extended radio structure is
likely related to typical radio lobes from the AGN, there may be some component of radio mini-halo emission.  A correlation
has been found between the presence of sloshing features in clusters and mini-halos (ZuHone, Markevitch, \& Brunetti 2011), and
these mini-halos may result from reacceleration of relativistic electrons by turbulence associated with sloshing (Mazzotta \& Giacintucci 2008).
With our current radio data, however, we do not see obvious correspondence between the radio emission and the spiral sloshing
feature.


There is evidence for outer bubbles related to one or more earlier outbursts from the AGN.  While not
seen in the surface brightness images, deficits are evident in a pressure-residual map after subtracting off a model
of the average pressure distribution.  The deficits are seen to the N and S of the cluster center, in line
with the current axis of the radio lobes.  Each of these bubbles could inject up to $1\times10^{60}$ erg into the intracluster medium.

\acknowledgements
Support for this work was provided by the National Aeronautics and Space
Administration, through {\it Chandra} Award Number
GO9-0147X.
Basic research in radio astronomy at the Naval Research Laboratory is supported by 6.1 Base funding.
SWR was supported in part by the Chandra X-ray Center through NASA contract NAS8-03060.
We thank Frazer Owen for useful discussions.

\clearpage

\begin{deluxetable}{ccccc}
\tabletypesize{\footnotesize}
\tablecaption{\it{{Chandra}} X-ray Observations}
\tablewidth{0pt}
\tablehead{
\colhead{Obs ID} & \colhead{Date} & \colhead{Roll Angle}
& \colhead{Data Mode} & \colhead{Exp}\\
\colhead{} & \colhead{} & \colhead{(deg)} & \colhead{} & \colhead{(ksec)}}
\startdata
890 &2000-09-03& 265.1 & FAINT  &  37.23   \\
5807 &2006-03-24& 98.9 & VFAINT &  128.63 \\
10477 &2009-06-05& 221.2 & VFAINT & 62.03   \\
10478 &2009-05-25 & 208.2 & VFAINT& 120.67   \\
10479 &2009-06-09 & 217.1 & VFAINT & 65.76    \\
10480 &2009-04-09 & 112.2 & VFAINT & 20.15     \\
10879 &2009-04-05 & 112.2 & VFAINT & 82.21     \\
10914 &2009-05-29 & 208.2 & VFAINT & 39.36    \\
10915 &2009-06-03 & 221.2 & VFAINT & 15.16   \\
10916 &2009-06-11 & 217.1 & VFAINT & 35.47   \\
10917 &2009-06-08 & 217.1 & VFAINT & 55.99   \\
\enddata
\end{deluxetable}

\begin{deluxetable}{lccccc}
\tabletypesize{\footnotesize}
\tablecaption{VLA Radio Observations}
\tablewidth{0pt}
\tablehead{
\colhead{Obs. Code} & \colhead{Date} & \colhead{VLA Configuration}   & \colhead{Frequency}   &
\colhead{Bandwidth} &
\colhead{Duration} \\
\colhead{} & \colhead{} & \colhead{} & \colhead{(MHz)} & \colhead{(MHz)} & \colhead{(hours)} 
}
\startdata
AS355 &1988-11-30 & A & 4835.1/4885.1 & 50/50 & 3.0 \\
AS355 &1989-03-30 & B & 4835.1/4885.1 & 50/50 & 2.4 \\
AS355 &1989-06-26 & C & 4835.1/4885.1 & 50/50 & 2.6 \\
AS355 &1988-11-30 & A & 1464.9/1514.9 & 50/50 & 3.0 \\
AS355 &1989-03-30 & A & 1464.9/1514.9 & 50/50 & 2.1 \\
\enddata
\label{tbl:radio}
\end{deluxetable}

\clearpage


\clearpage

\end{document}